\documentclass[9pt,twoside]{pnas-new}

\templatetype{pnasmathematics} 

\let\vec\boldsymbol

\setboolean{displaywatermark}{false}

\newcommand{\bvec}[1]{\mathbf{#1}}

\newcommand{\vs}{\bvec s}
\newcommand{\vy}{\bvec y}
\newcommand{\vr}{\bvec r}

\newcommand{\hn}{\hat{\bvec n}}

\newcommand{\hr}{\hat{\bvec r}}

\newcommand{\tj}[6]{\begin{pmatrix} {#1} & {#2} & {#3}\\ {#4} & {#5} & {#6}\end{pmatrix}}

\definecolor{darkgreen}{RGB}{0,120,0}

\newcommand{\resub}[1]{#1}

\renewcommand{\L}{{\vec{\Lambda}}}
\renewcommand{\P}{\mathcal{P}}

\newcommand{\vl}{\vec{\ell}}
\newcommand{\vL}{\vec{L}}
\newcommand{\ket}[1]{\left|{#1}\right\rangle}

\newcommand{\braket}[2]{\left\langle\left.{#1}\right|{#2}\right\rangle}
\newcommand{\cg}[3]{\left\langle\left.{#1},{#2}\right|{#3}\right\rangle}

\usepackage{empheq}

\usepackage{subcaption}

\def\beq{\begin{eqnarray}}
\def\eeq{\end{eqnarray}}

\title{Efficient Computation of $N$-Point Correlation Functions in $D$ Dimensions}

\author[a,b,1]{Oliver H.\,E.\,Philcox}
\author[c,d]{Zachary Slepian} 

\affil[a]{Department of Astrophysical Sciences, Princeton University, Princeton, NJ 08540, USA}
\affil[b]{School of Natural Sciences, Institute for Advanced Study, 1 Einstein Drive, Princeton, NJ 08540, USA}
\affil[c]{Department of Astronomy, University of Florida, 211 Bryant Space Science Center, Gainesville, FL 32611, USA}
\affil[d]{Physics Division, Lawrence Berkeley National Laboratory, 1 Cyclotron Road, Berkeley, CA 94709, USA}

\leadauthor{Philcox} 

\significancestatement{Stochastic processes appear throughout the physical sciences, and their properties are usually described by correlation functions. For discrete data, the $N$-point correlation function (NPCF) encodes the distribution of $N$-tuplets of points in space; \resub{estimation of the NPCF basis coefficients} from a set of $n$ particles scales as $n^N$. As $N$ increases, \resub{this measurement} becomes prohibitively expensive, thus statistics with $N>3$ are rarely used. Here, we show that NPCF \resub{components} may be estimated in $n^2$ time, by first expanding the statistics in separable angular bases. This approach has already found substantial application in quantifying galaxy clustering; here, we show it to be applicable to any homogeneous and isotropic space, regardless of dimension, \resub{and provide a practical implementation in \textsc{Julia}}.}

\authorcontributions{OP prepared the manuscript and generalized to $D$ dimensions, ZS assisted with the manuscript and developed the basis functions and algorithm in 3D Euclidean space.}
\authordeclaration{We declare no competing interests.}
\correspondingauthor{\textsuperscript{1}E-mail: ohep2\@cantab.ac.uk}

\keywords{Correlation Functions $|$ Clustering Statistics $|$ Computational Physics $|$ Cosmology $|$ Spherical Harmonics} 

\begin{abstract}
We present efficient algorithms for computing the $N$-point correlation functions (NPCFs) of random fields in arbitrary $D$-dimensional homogeneous and isotropic spaces. Such statistics appear throughout the physical sciences, and provide a natural tool to describe stochastic processes. Typically, \resub{algorithms for computing the NPCF components have} $\mathcal{O}(n^N)$ complexity (for a data set containing $n$ particles); their application is thus computationally infeasible unless $N$ is small. By projecting \resub{the statistic} onto a suitably-defined angular basis, we show that the estimators can be written in a separable form, with complexity $\mathcal{O}(n^2)$, or $\mathcal{O}(n_{\rm g}\log n_{\rm g})$ if evaluated using a Fast Fourier Transform on a grid of size $n_{\rm g}$. Our decomposition is built upon the $D$-dimensional hyperspherical harmonics; these form a complete basis on the $(D-1)$-sphere and are intrinsically related to angular momentum operators. Concatenation of $(N-1)$ such harmonics gives states of definite combined angular momentum, forming a natural separable basis for the NPCF. \resub{As $N$ and $D$ grow, the number of basis components quickly becomes large, providing a practical limitation to this (and all other) approaches: however, the dimensionality is greatly reduced in the presence of symmetries; for example, isotropic correlation functions require only states of zero combined angular momentum}. \resub{We provide a \textsc{Julia} package implementing our estimators, and show how they can be applied to a variety of scenarios within cosmology and fluid dynamics.} The efficiency of such estimators will allow higher-order correlators to become a standard tool in the analysis of random fields.  
\end{abstract}

\dates{This manuscript was compiled on \today}
\doi{\url{www.pnas.org/cgi/doi/10.1073/pnas.XXXXXXXXXX}}

\begin{document}

\maketitle
\thispagestyle{firststyle}
\ifthenelse{\boolean{shortarticle}}{\ifthenelse{\boolean{singlecolumn}}{\abscontentformatted}{\abscontent}}{}


\dropcap{R}andom fields are ubiquitous in the physical sciences. Perhaps the most powerful tool for their analysis is the set of \textit{$N$-point correlation functions} (hereafter NPCFs), defined as statistical averages over $N$ copies of the field at different spatial or temporal locations. If the field is Gaussian-random, only the first two correlators (the mean and two-point function) contain useful information, though this assumption is rarely true in practice. Examples of higher-order NPCFs populate many fields of study; a brief search will reveal their use in molecular physics \citep{OH-3pcf}, materials science \citep{berryman1988}, field theory \citep{dotsenko1991}, diffusive systems \citep{hwang1993,sanda2005} and cosmology \citep{2001misk.conf...71M}, amongst other topics. 

Correlation functions have found extensive use in the analysis of \resub{spectroscopic} galaxy surveys \citep[e.g.,][]{1978IAUS...79..217P}. Whilst the majority of information is contained within the 2PCF, inclusion of the higher-order functions is expected to significantly tighten constraints on cosmological parameters, particularly those pertaining to phenomena such as 
extensions to General Relativity \citep{2020arXiv201105771A}. Despite this, statistics beyond the 2PCF have been scarcely used in practice; in fact, almost no modern analyses have included correlators with $N>3$. The reason is simple: higher-order NPCFs are expensive to compute and analyze.

Consider a $D$-dimensional \resub{space} $\mathbb{M}^D$ (e.g., a Euclidean space) with an associated complex-valued random field $X:\mathbb{M}^D\to\mathbb{C}$. The NPCF, $\zeta:\left(\mathbb{M}^D\right)^{\otimes (N-1)}\to\mathbb{C}$ (for tensor product $\otimes$), is formally defined as
\beq\label{eq: zeta-def}
    \boxed{\zeta(\vr^1,\ldots,\vr^{N-1};\vs) = \mathbb{E}_X\left[X(\vs)X(\vs+\vr^1)\cdots X(\vs+\vr^{N-1})\right],}
\eeq
where $\mathbb{E}_X$ represents the statistical average over realizations of $X$, $\vs$ and $\vr^i$ are absolute and relative positions on the manifold, and we have assumed $N\geq 2$.
If the random field is \textit{statistically homogeneous}, all correlators must be independent of the absolute position $\vs$; this leads to the well-known NPCF estimator
\beq\label{eq: NPCF-estimator-full}
    \boxed{\hat\zeta(\vr^1,\ldots,\vr^{N-1}) = \frac{1}{V_D}\int_{\mathbb{M}^D} d^N\vs\,\left[X(\vs)X(\vs+\vr^1)\cdots X(\vs+\vr^{N-1})\right],}
\eeq
averaging $\vs$ over a volume $V_D$, and noting that the NPCF depends on only $(N-1)$ positions. \resub{In praxis, we cannot estimate the continuous NPCF of \eqref{eq: zeta-def}; rather, we consider only the quantity projected onto some basis functions, $B(\vr^1,\ldots,\vr^{N-1})$, for example a set of radial and angular bins. In this case, the problem reduces to estimating the coefficients $\zeta_B \equiv \int d\vr^1\cdots d\vr^{N-1}\zeta(\vr^1,\ldots,\vr^{N-1})B(\vr^1,\ldots,\vr^{N-1})$. 
Strictly, an infinite set of basis functions is required to fully specify the NPCF: conventionally, one restricts to a finite number, and applies the same basis projection to both data and theory, eliminating bias.}

\resub{The computational difficulties become clear if one considers measuring the NPCF components from a discrete field containing $n$ particles. In this instance, the random field $X$ can be represented as a sum over $n$ Dirac delta functions, \textit{i.e.} $X(\vr) = \sum_{j=1}^n w^j\delta^{\rm D}(\vr-\vy^j)$, where $\{\vy^j\}$ are the particle positions and $\{w^j\}$ are weights. Projecting onto a basis function $B$, the discrete version of \eqref{eq: NPCF-estimator-full} becomes}
\beq\label{eq: NPCF-estimator-intro-discrete}
    \resub{\hat\zeta_B = \frac{1}{V_D}\sum_{j_0=1}^nw^{j_0}\sum_{j_1=1}^nw^{j_1}\cdots\sum_{j_{N-1}=1}^n w^{j_{N-1}}B(\vy^{j_1}-\vy^{j_0},\ldots,\vy^{j_{N-1}}-\vy^{j_0}).}
\eeq
\resub{This is a sum over $N$-tuplets of particles, and, since the total number of $N$-tuplets scales as $n^N$, \textbf{has complexity $\mathcal{O}(n^N)$}} (cf.\,\S\ref{sec: NPCF-estimator}\ref{subsec: discrete-estimator}). Unless $n$ is very small, direct application of \eqref{eq: NPCF-estimator-intro-discrete} is infeasible for all but the smallest $N$.\footnote{See \citep{2005NewA...10..569Z} for an efficient, \resub{but inexact}, tree-based approach in 3D, \resub{as well as \citep{2019ApJS..242...29S} for a solution involving graph databases}.} 

\citep{2015MNRAS.454.4142S} presented a new technique to measure the \resub{3D Euclidean} 3PCF more efficiently, building on \citep{2004ApJ...605L..89S}. By \resub{representing $\zeta(\vr^1,\vr^2)$  in a factorizable angular basis of Legendre polynomials}, the former work obtained an algorithm \resub{for estimating 3PCF coefficients} with $\mathcal{O}(n^2)$ complexity. This algorithm facilitated a number of analyses, both in cosmology \citep[e.g.,][]{2017MNRAS.469.1738S,2017MNRAS.468.1070S} and magnetohydrodynamics \citep{2018ApJ...862..119P}. Our companion paper \citep{npcf_algo} showed that the approach can be generalized to \resub{the computation of rotationally invariant NPCF coefficients} in 3D Euclidean space, which is of particular relevance for galaxy surveys. 
Here, we show  that similar estimators may be constructed in \resub{any} homogeneous and isotropic space; in particular, \resub{if the basis is carefully chosen,} \textbf{an $\mathcal{O}(n^2)$ estimator \resub{for each basis component} is possible, regardless of $N$, $D$, and the spatial curvature.} \resub{Furthermore, if the data can be mapped to a grid of dimension $n_g$, Fast Fourier Transforms (FFTs) can be used to reduce this to $\mathcal{O}(n_g\log n_g)$.}

Our pathway to obtaining an efficient NPCF estimator is the following:
\begin{enumerate}
    \item Construct a set of basis functions for the angular part of a homogeneous and isotropic space $\mathbb{M}^D$ (\S\ref{sec: single-basis}). \resub{A natural choice is the set of} \textit{hyperspherical harmonics}, which \resub{generalize} the spherical harmonics and arise in the $D$-dimensional theory of angular momentum.
    \item Combine $(N-1)$ hyperspherical harmonics to create an $(N-1)$-particle \resub{angular} basis on the manifold $\mathbb{M}^D$ (\S\ref{sec: npoint-basis}). In particular, we form states of definite combined angular momentum, denoted $\P_{\vec \Lambda}^{\vL}$. \resub{Since the (statistically homogeneous) NPCF depends only on $(N-1)$ coordinates, its angular part can be decomposed into this basis. Explicitly:}
    \beq\label{eq: zeta-basis-decomp-intro}
    \resub{\zeta(\vr^1,\ldots,\vr^{N-1}) = \sum_{\vL}\sum_{\L} \zeta^{\vL}_\L(r^1,\ldots,r^{N-1})\P_{\L}^{\vec L}(\hr^1,\ldots,\hr^{N-1}),}
    \eeq
    \resub{where $\zeta^{\vL}_{\L}$ are the basis components, the sets of indices $\L$ and $\vL$ represent internal and external angles of the NPCF, $\hr$ represents the unit vector parallel to $\vr$, and $r \equiv |\vr|$. In the language of \eqref{eq: NPCF-estimator-intro-discrete}, our basis functions are $B(\vr^1,\ldots,\vr^{N-1}) = \Theta^{\vec b}(r^1,\ldots,r^{N-1})\P_{\L}^{\vec L}(\hr^1,\ldots,\hr^{N-1})$, where $\Theta^{\vec b}$ is some separable set of radial bins indexed by the vector $\vec b$ (see Eq.\,\ref{eq: B-hypersph}).}
    \item \resub{Using \eqref{eq: NPCF-estimator-full}, construct an estimator for the NPCF basis components $\zeta_{\L}^{\vL}$ (\S\ref{sec: NPCF-estimator}).} Since the basis functions are separable in $\vr^1,\vr^2,\ldots$, \resub{the estimator can be factorized, and takes the schematic form} (cf.\,Eq.\,\ref{eq: NPCF-estimator-discrete})
    \beq\label{eq: efficient-estimator-sketch}
        \resub{\hat\zeta_{\L}^{\vL}(r^1,\ldots,r^{N-1}) = \frac{1}{V_D}\sum_{j=1}^n w^j \times \sum_{\vl^1\cdots\vl^{N-1}} \left[\mathrm{coupling}\right] \times a_{\vl^1}(\vy^j;r^1)\cdots a_{\vl^{N-1}}(\vy^j;r^{N-1}),}
    \eeq
    \resub{where $\vl$ specify angular momentum indices, and each $a_{\vl}(\vy;r)$ function involves a further sum over $n$ particles, weighted by a hyperspherical harmonic (see Eq.\,\ref{eq: NPCF-estimator-discrete}). Since each $a_{\vl}(\vs;r)$ component can be estimated independently,} the algorithm has complexity $\mathcal{O}(n^2)$, \resub{or $\mathcal{O}(n_g\log n_g)$ using an FFT with $n_g$ grid points}. \resub{\eqref{eq: efficient-estimator-sketch} can be applied to each basis component separately; we caution that the total number of components (for a fixed angular and radial resolution) becomes large as $N$ and $D$ increase, providing a practical limitation to any NPCF algorithm.}
\end{enumerate}
\resub{In \S\ref{sec: applications} we discuss a numerical implementation of the NPCF estimator,}\footnote{\resub{Available at \href{https://github.com/oliverphilcox/NPCFs.jl}{GitHub.com/oliverphilcox/NPCFs.jl}.}} alongside a variety of applications.





\begin{figure}[t]
\centering
\begin{minipage}{.45\textwidth}
    \centering
    \vskip 0.6cm
    \includegraphics[width=\textwidth]{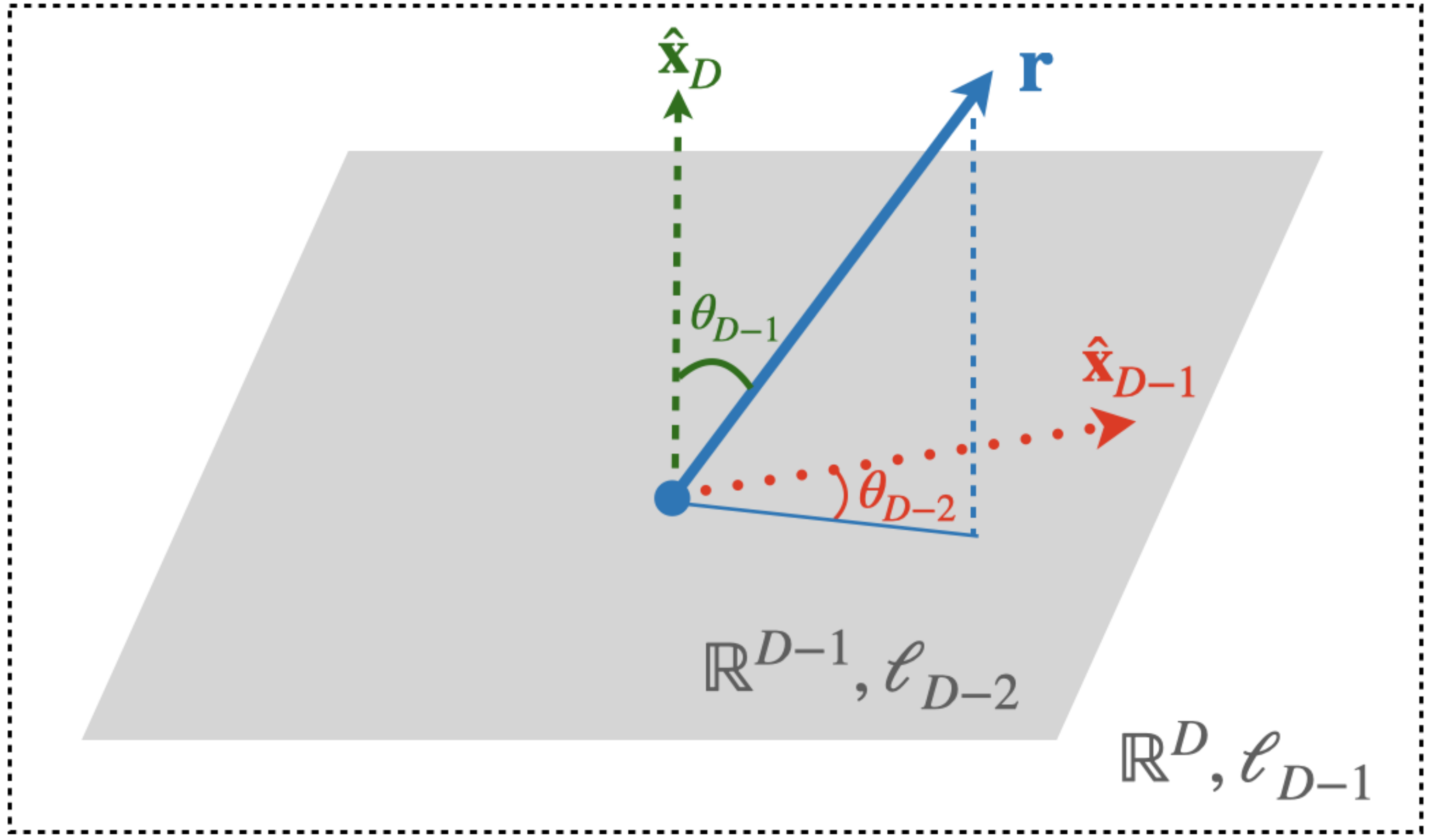}
    \caption{Cartoon illustrating the coordinate system used in this work, assuming a Euclidean geometry. The outer dashed box indicates $\mathbb{R}^D$, whilst the gray plane represents $\mathbb{R}^{D-1}$, containing the Cartesian coordinates $\{x_1,\ldots,x_{D-1}\}$. The position vector $\vr\in\mathbb{R}^D$ can be expressed in hyperspherical coordinates $\{r,\theta_1,\ldots,\theta_{D-1}\}$, where $r$ is a radial coordinate. $\theta_{K-1}$ is defined as angle between the \resub{$\hat{\bvec x}_{K}$} axis and the projection of $\vr$ into the subspace $\mathbb{R}^{K}$, as shown for $\theta_{D-1}$ and $\theta_{D-2}$. The angles obey the restrictions $\theta_K\in[0,\pi)$ for $K>1$, and $\theta_1\in[0,2\pi)$. \resub{Each angle can be associated with an angular momentum index $\ell_{K-1}$, encoding} the orbital angular momentum in the subspace $\mathbb{R}^K$ \resub{in which only the first $K$ Cartesian coordinates are varied} (\S\ref{sec: single-basis}). $\ell_1$ gives the azimuthal angular momentum \resub{in the subspace containing $\{x_1,x_2,x_3\}$}, and $\ell_{D-1}$ gives the total angular momentum.}
    \label{fig: coords}
\end{minipage}%
\quad
\begin{minipage}{.44\textwidth}
    \centering
    \includegraphics[width=\textwidth]{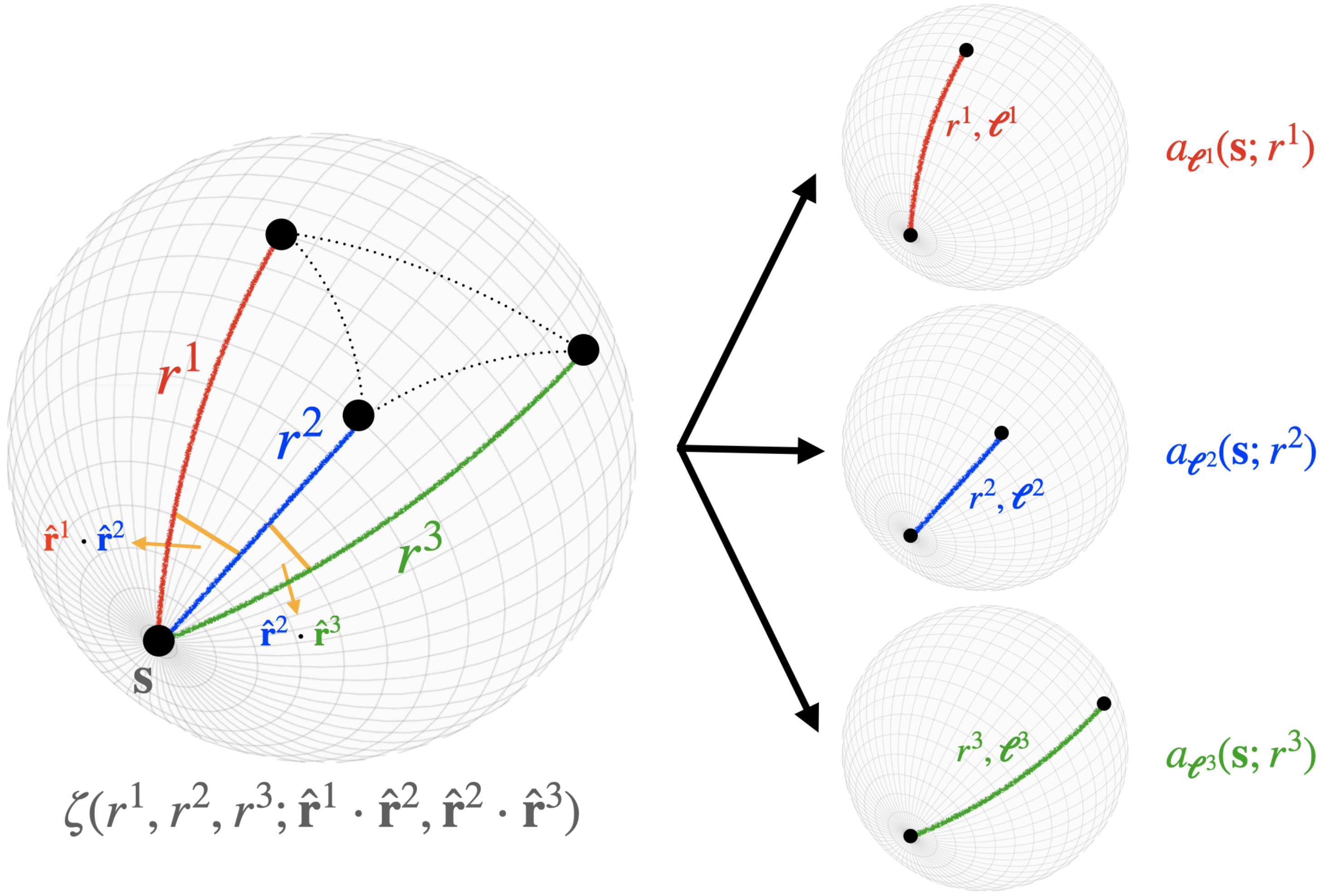}
    \caption{Sketch of the decomposition underlying our NPCF algorithm, visualized for $N=4$ in a \resub{2D spherical geometry}. On the left, we show the \resub{(rotation-averaged)} 4PCF defined by three \resub{distances, $r^1$, $r^2$, $r^3$, and two angles $\hr^1\cdot\hr^2$, $\hr^2\cdot\hr^3$}, relative to a primary position $\vs$; na\"ive 4PCF estimation from $n$ particles proceeds by \resub{summing over each of the possible $n^4$ sets of points}. On the right, we show our decomposition, factorizing the 4PCF into three functions of two positions, $a_{\vl}(\vs;r)$, which may be independently estimated from the data-set, \resub{requiring consideration of only $n^2$ pairs of particles}. 
    Each function depends on a side length $r^i$ and a set of angular momentum indices $\vl^i$; the latter specifies \resub{$\hr^i$} in the hyperspherical harmonic basis. This permits \resub{the NPCF components to be estimated by an algorithm} with $\mathcal{O}(n^2)$ complexity.}
        \label{fig: npcfs}
 \end{minipage}
 \end{figure}

\section{Single-Particle Basis}\label{sec: single-basis}
We begin by discussing the angular basis for functions of one position in $\mathbb{M}^D$, hereafter referred to as the `single-particle' basis. In \S\ref{sec: npoint-basis}, the basis will allow construction of a joint basis of $(N-1)$ positions onto which the NPCF can be projected.

\subsection{Constant-Curvature Metric}\label{subsec: metric}
To form an efficient angular basis, we require the underlying manifold to be (a) homogeneous, and (b) isotropic. This leads to the well-known line-element
\beq\label{eq: metric}
    \resub{d\Sigma^2_{\mathbb{M}^D}  =  dr^2+\chi_k^2(r)\left\{d\theta_{D-1}^2+\sin^2\theta_{D-1}\left[d\theta_{D-2}^2+\sin^2\theta_{D-2}\left(d\theta_{D-3}^2+\cdots\right)\right]\right\}}
\eeq
\citep[e.g.,][]{1990MNRAS.247...57C}, adopting hyperspherical coordinates $\vr\equiv\{r,\theta_1,\ldots,\theta_{D-1}\}$.\footnote{\resub{Note that some conventions label the $\theta$ coordinates in the opposite order.}} In this parametrization, $r$ is a radial coordinate, whilst the $\theta_i$ are angular variables, with $\theta_1\in[0,2\pi)$ (often denoted by $\phi$ in $D=3$) and $\theta_j\in[0,\pi)$ for $j>1$. A sketch of the coordinates is shown in Fig.\,\ref{fig: coords} for a Euclidean geometry. \eqref{eq: metric} is a \textit{constant-curvature} metric, specified by \resub{$\chi_k(r) = \sin\left(r\sqrt{k}\right)/\sqrt{k}$ if $|k|>0$ and $\chi_k(r) = r$ else.}
Here, $k>0$ gives the $D$-sphere, $\mathbb{S}^D$, $k<0$ leads to the hyperbolic geometry $\mathbb{H}^D$, and $k=0$ results in a Euclidean geometry $\mathbb{R}^D$. 
Such manifolds are ubiquitous in the physical sciences; for instance, this describes the spatial part of the Friedmann-Lema\^itre-Robertson-Walker metric for an expanding Universe \citep{1984ucp..book.....W}.

\subsection{\resub{Hyperspherical Harmonics}}\label{subsec: hypersphericals}
A convenient basis for the constant-curvature space $\mathbb{M}^D$ is formed from the set of \textit{harmonic} functions $H:\mathbb{M}^D\to\mathbb{C}$. These satisfy the \textit{Laplace-\resub{Beltrami}} equation
\beq\label{eq: laplace-bertrami-eq}
    \Delta H(\vr) \equiv \frac{1}{\sqrt{|g|}}\partial_i \left(\sqrt{|g|}\;g^{ij}\partial_j H(\vr)\right) = 0,
\eeq
where $H$ is twice continuously differentiable, $g$ is the metric (with $|g| = \mathrm{det}\left[g_{ij}\right]$), $i\in\{1,2,\ldots,D\}$ and we have assumed the Einstein summation convention. \resub{Assuming the metric of \eqref{eq: metric}, \eqref{eq: laplace-bertrami-eq} permits a separable solution of the form $H(\vr)=R(r)Y(\hr)$, where $\hr = \{\theta_1,\ldots,\theta_{D-1}\}$. The angular part of this must satisfy the following eigenfunction equation for constant $\lambda_{D-1}$:}
\beq\label{eq: eigenfunction-S-D-1}
    \Delta_{\mathbb{S}^{D-1}}Y(\hr) = -\lambda_{D-1} Y(\hr),
\eeq
where $\Delta_{\mathbb{S}^{D-1}}$ is the Laplace-\resub{Beltrami} operator on the $(D-1)$-sphere, given explicitly in \eqref{eq: eigenfunction-eqns}. The corresponding solutions are the \textit{hyperspherical harmonics} on $\mathbb{S}^{D-1}$ \citep[e.g.,][]{cooper1963,cohl2011,2012arXiv1205.3548F}. \resub{Since the harmonic functions are separable, the hyperspherical harmonics form an angular basis for \textit{any} function on $\mathbb{M}^{D}$, regardless of the spatial curvature $k$.}\footnote{This is additionally seen by noting that \eqref{eq: metric} can be written $d\Sigma^2_D = dr^2 + \chi_k^2(r)d\Omega_{D-1}^2$, where $d\Omega_{D-1}^2$ is the line element on the $(D-1)$-sphere.} Physically, this decomposition is guaranteed since we have assumed the metric to be homogeneous and isotropic, enforcing invariance under the rotation group $SO(D)$ about any origin.

\resub{For a given dimension $D$, the hyperspherical harmonics, denoted $Y_{\ell_1\ldots\ell_{D-1}}(\hr)$, may be obtained by solving \eqref{eq: eigenfunction-S-D-1} recursively, as detailed in Appendix \ref{appen: hyperspherical}. These depend on a 
set of $(D-1)$ integers, $\{\ell_k\}$, which are related to angular momentum (cf.\,\S\ref{sec: single-basis}\ref{subsec: ang-mom-single}), and satisfy} the selection rules
\beq\label{eq: selection-rules}
    -\ell_2 \leq \ell_1 \leq \ell_2, \quad \ell_{K-1} \leq \ell_{K} \leq \ell_{K+1} \quad (2\leq K\leq D-1).
\eeq
\resub{In two- and three-dimensions, the hyperspherical harmonics take a simple form:}
\beq
    \resub{Y_{\ell_1}(\theta_1) = \frac{1}{\sqrt{2\pi}}\,e^{i\ell_1\theta_1}, \qquad Y_{\ell_1\ell_2}(\theta_1,\theta_2) = \sqrt{\frac{2\ell_2+1}{4\pi}\frac{(\ell_2+\ell_1)!}{(\ell_2-\ell_1)!}}\,e^{i\ell_1\theta_1}P_{\ell_2}^{-\ell_1}(\cos\theta_2),}
\eeq
\resub{for associated Legendre polynomials $P_\ell^{m}$. The $D=3$ functions are the usual spherical harmonics (in the Condon-Shortley convention), made clear by the identification $(\ell_1,\ell_2)\to(m,\ell)$, $(\theta_1,\theta_2)\to(\phi,\theta)$, where $\phi$ is the azimuthal angle. The explicit form of the hyperspherical harmonics for general $D$ is given in \eqref{eq: hypersph-def}.}

\subsection{Connection to Angular Momentum Eigenstates}\label{subsec: ang-mom-single} 

In 3D, the theory of angular momentum is centered around two operators, $\widehat{L}^2$ and $\widehat{L}_3$, which respectively give the total angular momentum and that projected onto the $\hat{\bvec x}_3$ axis. Both may be constructed from the vector operator $\widehat{\bvec L} = \bvec r \times \bvec p$, where $p_j = -i(\partial/\partial x^j)$ is the linear momentum in $\mathbb{R}^3$. For $D\neq 3$, we cannot define a cross-product, thus we instead start from the \textit{tensorial} 
angular momentum operator, following \citep{louck1960}:
\beq\label{eq: ang-mom-operator}
    \widehat{\mathsf L}_{ij} = x_ip_j - x_jp_i, \quad i\neq j \in \{1,2,\ldots, D\}
\eeq
where $\{x_1,\ldots,x_D\}$ is a Cartesian coordinate chart.\footnote{Since $\mathbb{M}^D$ and $\mathbb{R}^D$ share the same angular parametrization, and angular momentum is independent of radial coordinates, we may work in a Euclidean space for the purposes of this section.} These operators are antisymmetric ($\widehat{\mathsf L}_{ij} = -\widehat{\mathsf L}_{ji}$) and form the Lie algebra $\mathfrak{so}(D)$, \textit{i.e.} \resub{that of the rotation group in $D$ dimensions}. Of particular interest is $\widehat{\mathsf L}_{12}$; when applied to some state, this gives the \resub{azimuthal angular momentum in the subspace containing $\{x_1,x_2,x_3\}$}, just as for $\widehat{L}_3$ in $D=3$.

To fully define the rotational properties of a \resub{single-particle function in $D$-dimensional space}, we must specify not only its total angular momentum, but also \resub{that} projected into lower-dimensional subspaces. \resub{These} can be obtained from the operators
\beq\label{eq: k-dim-ang-mom}
    \widehat{\mathcal L}^2_{K} \equiv \sum_{i=1}^{K}\sum_{j=i+1}^K(\widehat{\mathsf L}_{ij})^2,
\eeq
for $K\geq 3$. Here, $\widehat{\mathcal L}^2_3 \equiv (\widehat{\mathsf L}_{12})^2+(\widehat{\mathsf L}_{13})^2+(\widehat{\mathsf L}_{23})^2$ gives the angular momentum in the subspace \resub{$\{x_1,x_2,x_3,0,\ldots,0\}$ of} $\mathbb{R}^D$, $\widehat{\mathcal L}^2_4$ gives that in the subspace \resub{$\{x_1,x_2,x_3,x_4,0,\ldots,0\}$} \textit{et cetera.} $\widehat{\mathcal L}_{D}^2$ is the \resub{total} angular momentum operator, analogous to $\widehat{L}^2$ in the three-dimensional theory. 

As shown in \citep[][Ch.\,3]{louck1960}, a complete set of commuting angular momentum operators is given by $\{\widehat{\mathsf{L}}_{12}$, $\widehat{\mathcal{L}}_3^2, \ldots, \widehat{\mathcal{L}}_{D}^2\}$. Moreover, the hyperspherical harmonics of \S\ref{sec: single-basis}\ref{subsec: hypersphericals} are simultaneous eigenfunctions of these, satisfying\footnote{This occurs since the $K$-dimensional Laplace-\resub{Beltrami} operator of \eqref{eq: eigenfunction-eqns} (which the hyperspherical harmonics are eigenfunctions of) is related to the $K$-dimensional total angular momentum operator by $\Delta_{\mathbb{S}^{K-1}} = -\widehat{\mathcal L}_K^2$.}
\beq\label{eq: single-particle-eigenfun}
    \widehat{\mathsf L}_{12} Y_{\ell_1\ldots\ell_{D-1}}(\hr) &=& \ell_1\,Y_{\ell_1\ldots\ell_{D-1}}(\hr), \qquad \widehat{\mathcal L}^2_{K} Y_{\ell_1\ldots\ell_{D-1}}(\hr) = \ell_{K-1}(\ell_{K-1}+K-2)\,Y_{\ell_1\ldots\ell_{D-1}}(\hr) \quad (3\leq K\leq D).
\eeq
We may thus associate $\ell_{K-1}$ with the orbital angular momentum in the subspace \resub{comprising} the first $K$ Cartesian coordinates (for $K\geq 3)$, and $\ell_1$ with that projected onto the $\hat{\bvec x}_3$ axis. Such an interpretation also justifies the conditions in \eqref{eq: selection-rules}; projections of the angular momentum into lower-dimensional spaces must have equal or \resub{lesser} magnitudes than the total angular momentum in $D$ dimensions. 

For later use, we introduce abstract notation for the angular momentum basis functions, \resub{writing $Y_{\ell_1\ldots\ell_{D-1}}(\hr) \to \ket{\ell_1\ldots\ell_{D-1}}$ in Dirac (or bra-ket) notation.}\footnote{The Hilbert space formed from these states is an infinite-dimensional representation of the rotation group $SO(D)$. Practically, the eigenvalue $\ell_{D-1}$ represents the behavior in $SO(D)$, whilst the values of $\ell_{K-1}$ with $K<D$ specify the rotation's action in a lower-dimensional subgroup $SO(K)$.} In total, there are $\binom{\ell_{D-1}+D-3}{D-3}(2\ell_{D-1}+D-2)/(D-2)$ eigenstates corresponding to a given total angular momentum $\ell_{D-1}$ if $D\geq 3$, and one if $D=2$ \citep[Eq.\,3.37]{louck1960}. \resub{For a suitably defined inner product, the basis functions are orthonormal} \citep{higuchi1987,cohl2011,2012arXiv1205.3548F}:
\beq
    \int_{\mathbb{S}^{D-1}}d\Omega_{D-1}Y^*_{\ell_1^{}\ldots\ell^{}_{D-1}}(\hr)Y_{\ell_1'\ldots\ell'_{D-1}}(\hr) \equiv \braket{\ell_1^{}\ldots\ell^{}_{D-1}}{\ell_1'\ldots\ell'_{D-1}} = \delta^{\rm K}_{\ell_1^{}\ell_1'}\ldots \delta^{\rm K}_{\ell_{D-1}^{}\ell_{D-1}'},
\eeq
where the Kronecker delta, $\delta^{\rm K}_{ij}$, is unity if $i=j$ and zero otherwise. \resub{Furthermore, they form} a complete basis on $\mathbb{S}^{D-1}$, such that, for any $h:\mathbb{S}^{D-1}\to\mathbb{C}$
\beq\label{eq: 1d-basis-decomp}
     \ket{h} = \sum_{\ell_1\ldots\ell_{D-1}}\braket{\ell_1\ldots\ell_{D-1}}{h}\ket{\ell_1\ldots\ell_{D-1}} \quad \leftrightarrow \quad ,h(\hr) = \sum_{\ell_1\ldots\ell_{D-1}}h_{\ell_1\ldots\ell_{D-1}}Y_{\ell_1\ldots\ell_{D-1}}(\hr)
\eeq
where the summation \resub{runs} over all angular momentum indices allowed by the selection rules of \eqref{eq: selection-rules}. \resub{The} basis coefficients $h_{\ell_1\ldots\ell_{D-1}}=\braket{\ell_1\ldots\ell_{D-1}}{h}$ may be obtained via orthonormality. Since the angular part of $\mathbb{M}^D$ is just the $(D-1)$-sphere $\mathbb{S}^{D-1}$ (\S\ref{sec: single-basis}\ref{subsec: metric}), \textbf{any one-particle function on $\mathbb{M}^D$ can be decomposed into this basis}; in general, the coefficients retain dependence on the radial coordinate $r$.


\section{$(N-1)$-Particle Basis}\label{sec: npoint-basis}
We now utilize the mathematics of angular momentum addition to generalize the angular basis of \S\ref{sec: single-basis} to functions of $(N-1)$ positions, \textit{i.e.} $f:\left(\mathbb{M}^D\right)^{\otimes (N-1)}\to\mathbb{C}$.

\subsection{Angular Momentum Addition}\label{subsec: ang-mom-add}
To begin, we consider the combination of two angular basis functions on the $(D-1)$-sphere. For convenience, we will work in the Dirac representation and denote the set of angular momentum indices by $\vl\equiv\{\ell_1,\ldots,\ell_{D-1}\}$, with superscripts used to distinguish between particles. Given single-particle states $\ket{\vl^1}$ and $\ket{\vl^2}$, the simplest two-particle state is $\ket{\vl^1,\vl^2} \equiv \ket{\vl^1}\ket{\vl^2}$, which exists in the product space $\mathbb{S}^{D-1}\otimes\mathbb{S}^{D-1}$. This is a simultaneous eigenstate of angular momentum operators for the first and second particles, $\widehat{\mathsf L}_{ij}^{(1)}$ and $\widehat{\mathsf L}_{ij}^{(2)}$ 
(cf.\,\S\ref{sec: single-basis}\ref{subsec: ang-mom-single}).
Whilst the product states $\ket{\vl^1,\vl^2}$ do form a basis on $\mathbb{S}^{D-1}\otimes\mathbb{S}^{D-1}$ (sometimes called the `uncoupled basis' \citep{2020arXiv200205011T}), it is not an efficient one, since (a) we require $2(D-1)$ angular momentum indices to specify the state, which, as shown below, is considerably more than necessary, and (b) the indices are not straightforwardly connected to the \resub{joint} rotation properties of the two-particle state.

A more appropriate basis is wrought by considering the \resub{combined} angular momentum operator $\widehat{\mathsf L}_{ij}^{(12)} \equiv \widehat{\mathsf L}_{ij}^{(1)} + \widehat{\mathsf L}_{ij}^{(2)}$, which specifies the properties of some two-particle function, $f(\vr^1,\vr^2)$, under joint rotations of $\vr^1$ and $\vr^2$ about a common origin. As in \S\ref{sec: single-basis}\ref{subsec: ang-mom-single}, $\widehat{\mathsf L}_{ij}^{(12)}$ may be used to construct a set of commuting angular momentum operators, whose eigenstates can be written $\ket{\vL}\equiv\ket{L_1\ldots L_{D-1}}$. Here, $L_{K-1}$ specifies the \resub{combined} angular momentum projected into the $K$-dimensional subspace \resub{in which the first $K$ Cartesian coordinates are varied} (for $K\geq 3$), and $L_1$ gives that projected onto the $\hat{\bvec x}_3^1,\hat{\bvec x}_3^2$ axes. Similarly to the single-particle state, the $\vL$ indices must obey the selection rules of \eqref{eq: selection-rules}.

Two-particle states of definite combined angular momentum are formed by summing over products of one-particle states, just as in the 3D case (e.g.,\,\citep{biedenharn_louck_carruthers_1984,1988qtam.book.....V}, see also \citep{2014arXiv1412.5829S,2019PhRvC..99c4320T,2020arXiv200205011T} for a \resub{discussion with more general $D$}). \resub{Explicitly, they are given by}
\beq\label{eq: ang-mom-add-N=3}
    \boxed{\ket{\ell_{D-1}^{1}\;\ell_{D-1}^2;\vL} = \sum_{\ell_1^{1}\ldots\ell^{1}_{D-2}}\sum_{\ell_2^2\ldots\ell^2_{D-2}}\cg{\vl^1}{\vl^2}{\vL} \ket{\vl^1}\ket{\vl^2},}
\eeq
where $\cg{\vl^1}{\vl^2}{\vL}$ is a Clebsch-Gordan (hereafter CG) coefficient \citep[e.g.,][]{VanIsacker1991}. This is often referred to as the `coupled basis' \citep{2020arXiv200205011T}. To uniquely define the state, we must specify (a) the combined angular momentum eigenvalues $\vL$, and (b) the total angular momentum of the first and second particle, $\ell_{D-1}^1$ and $\ell_{D-1}^2$. Importantly, \eqref{eq: ang-mom-add-N=3} involves a sum over $\{\ell_1^i,\ldots,\ell_{D-2}^i\}$, \textit{i.e.} the projection of the particles' angular momentum into lower-dimensional subspaces. Due to this, the combined state is specified by only $(D+1)$ indices, which, for $D>3$, is significantly fewer than the $2(D-1)$ required for $\ket{\vl^1,\vl^2}$.

Practically, one must know the CG coefficients $\cg{\vl^1}{\vl^2}{\vL}$ in order to form the \resub{coupled basis} of \eqref{eq: ang-mom-add-N=3}.\footnote{The CG coefficients used in this work are those of the $SO(D)\supset SO(D-1)\supset \cdots\supset SO(2)$ reduction.} These have been studied in depth \citep[e.g.,][]{biedenharn1961,biedenharn_louck_carruthers_1984,1988qtam.book.....V,VanIsacker1991,Louck2006,2010JMP....51i3518C} and may be simplified by techniques such as Racah's factorization lemma \citep{Racah1965}. One route to their computation is via \eqref{eq: ang-mom-add-N=3}; starting from a state of maximal combined angular momentum, $\ket{\ell_{D-1}^{1}\ell_{D-1}^2;\vl^1+\vl^2} \equiv \ket{\vl^1 }\ket{\vl^2}$ (wherein the two angular momentum vectors in $\mathbb{R}^{D}$ are aligned), ladder operators may be applied iteratively to obtain states of lower angular momentum, whose weightings give the CG coefficients. \resub{As in \citep{1988qtam.book.....V,biedenharn_louck_carruthers_1984}, the explicit forms for the CG coefficients in $D=2$ and $D=3$ are given by}
\beq\label{eq: CG-examples}
    \resub{\cg{\ell^1}{\ell^2}{L} = \delta^{\rm K}_{(\ell^1+\ell^2)L}, \qquad \cg{\ell^1_1\ell^1_2}{\ell_1^2\ell_2^2}{L_1L_2} = (-1)^{-\ell_2^1+\ell_2^2-L_1}\sqrt{2L_2+1} \tj{\ell_2^1}{\ell_2^2}{L_2}{\ell_1^1}{\ell_1^2}{-L_1},}
\eeq
\resub{where the $2\times 3$ matrix is a Wigner 3-$j$ symbol \citep[e.g.,][\S34]{NIST:DLMF}. The $D=4$ case is similar \citep[see][]{biedenharn1961,VanIsacker1991,2006JPhA...39.3099M}, but includes a Wigner 9-$j$ symbol.}


CG coefficients satisfy certain orthogonality conditions, including
\beq\label{eq: CG-orthonorm}
    \sum_{\ell^{1}_1\ldots\ell_{D-2}^{1}}\sum_{\ell^2_1\ldots\ell^2_{D-2}}\cg{\vl^1}{\vl^2}{\vL}\cg{\vl^1}{\vl^2}{\vL'} = \delta^{\rm K}_{L_1^{}L_1'}\times \cdots\times \delta^{\rm K}_{L_{D-1}^{}L_{D-1}'}.
\eeq
Coupled with the orthonormality of the one-particle states $\ket{\vl}$ (Eq.\,\ref{eq: N-1-particle-orthonormality}), this ensures that the combined-angular-momentum basis is orthonormal, \textit{i.e.}
\beq
    \braket{\resub{l^1_{D-1} l^2_{D-1};\vL'}}{\ell^1_{D-1}\ell_{D-1}^2;\vL} = \left(\delta^{\rm K}_{L^{}_1L_1'}\cdots\delta^{\rm K}_{L^{}_{D-1}L'_{D-1}}\right)\times\left(\delta^{\rm K}_{\ell^1_{D-1}l^1_{D-1}}\delta^{\rm K}_{\ell_{D-1}^2l_{D-1}^2}\right).
\eeq
Furthermore, $\cg{\vl^1}{\vl^2}{\vL}$ is non-zero only if the \resub{following conditions} are satisfied:
\beq\label{eq: addition-triangle-conditions}
    L_1 = \ell^{1}_1 + \ell^2_1, \quad |\ell^{1}_K-\ell^2_K|\leq L_K \leq \ell^{1}_K+\ell^2_K, \quad \resub{(2\leq K\leq D-1)}.
\eeq
The first constraint occurs since $L_1$ is the eigenvalue corresponding to $\widehat{\mathsf L}^{(12)}_{12}$, which is linear in $\widehat{\mathsf L}_{12}^{(1)}$ and $\widehat{\mathsf L}_{12}^{(2)}$ (cf.\,addition of $m$ indices in 3D), and the second is due to the triangle inequality, recalling that $L_K$ corresponds to the magnitude of the angular momentum in the subspace $\mathbb{R}^{K}$. In particular, the constraints fix the total combined angular momentum, $L_{D-1}$, to be no greater than $\ell_{D-1}^1+\ell_{D-1}^2$.

\subsection{Combined Angular Momentum Basis}
By repeated application of the angular momentum addition rule (Eq.\,\ref{eq: ang-mom-add-N=3}), we may build up an $(N-1)$-particle state of definite combined angular momentum. \resub{First, we combine $\ket{\vl^1}$ and $\ket{\vl^2}$ to form the two-particle state $\ket{\ell^1_{D-1}\ell^2_{D-1};\vl^{12}}$, which is then combined with $\vl^3$ to form the three-particle state $\ket{\ell^1_{D-1}\ell^2_{D-1}\ell^{12}_{D-1}\ell^{3}_{D-1};\vl^{123}}$, \textit{et cetera}.}
\resub{In full, we obtain the} $(N-1)$-particle basis function
\beq\label{eq: combined-ang-mom-states}
    \boxed{
    \ket{\L;\vL} = \sum_{[\vl^1][\vl^2]\ldots[\vl^{N-1}]}C^{\L;\vL}_{\vl^1\vl^2\ldots\vl^{N-1}}\ket{\vl^1}\ket{\vl^2}\cdots\ket{\vl^{N-1}},}
\eeq
summing over all intermediate angular momenta $\ell_{k}^i$ with $k<(D-1)$, as denoted by $[\vl^i]$.\footnote{For $D=3$, our basis functions match the coupled representation of $SU(2)$ discussed in \citep{2020arXiv200205011T} in the context of quantum chemistry.} 
\resub{The combined state in \eqref{eq: combined-ang-mom-states} is specified by} the set of total angular momentum indices $\L \equiv \{\ell^1_{D-1},\ell_{D-1}^2,\ell_{D-1}^{12},\ell^3_{D-1},\resub{\ell^{123}_{D-1},\ell^4_{D-1}},\ldots,\ell_{D-1}^{12\ldots(N-2)},\ell_{D-1}^{{N-1}}\}$ and \resub{involves the coupling coefficients}
\beq\label{eq: Coupling-C}
    C^{\L;\vL}_{\vl^1\vl^2\ldots\vl^{{N-1}}} = \sum_{[\vl^{12}]\ldots[\vl^{12\ldots(N-2)}]}\cg{\vl^1}{\vl^2}{\vl^{12}}\cg{\vl^{12}}{\vl^3}{\vl^{123}}\cdots\cg{\vl^{12\ldots(N-2)}}{\vl^{N-1}}{\vec L},
\eeq
\resub{which is a product of CG symbols.} We have additionally set $\vl^{12\ldots (N-1)}\equiv \vL$, representing the combined angular momentum eigenvalues. Note that \eqref{eq: combined-ang-mom-states} contains a sum over both the primary indices $\ell^{i}$ (which define the single-particle states) and intermediates, e.g., $\ell^{12\ldots}$ (within the coupling definitions), but not those corresponding to total angular momenta, \textit{i.e.} $\ell_{D-1}$. 

The meaning of \eqref{eq: combined-ang-mom-states} is straightforward; an $(N-1)$-particle state with \resub{combined angular momentum} $\vL$ can be obtained as a sum of product states, weighted by $(N-2)$ CG coefficients. To define the state uniquely, we must specify not only the total angular momentum $\ell_{D-1}$ of each single-particle state, but also the total angular momentum of the intermediate states, e.g., $\ell_{D-1}^{12}$ arising from the coupling of $\ket{\vl^1}$ and $\ket{\vl^2}$. In total, the state is specified by $(2N+D-5)$ indices; again significantly fewer than the $(N-1)(D-1)$ required for the product state $\ket{\vl^1}\cdots\ket{\vl^{N-1}}$.

A particularly interesting state is that of \resub{zero} combined angular momentum, \textit{i.e.} $\vL = \vec 0$.\footnote{For $\mathbb{M}^D = \mathbb{R}^3$, our treatment of the $\vL = \vec 0$ states exactly follows that of \citep{2020arXiv201014418C}.} From \eqref{eq: combined-ang-mom-states}, this is simply
\beq\label{eq: zero-ang-mom-states}
    \boxed{\ket{\L;\vec{0}} = \sum_{[\vl^1]\ldots[\vl^{N-1}]}C^{\L;\vec 0}_{\vl^1\ldots\vl^{{N-1}}}\ket{\vl^1}\ket{\vl^2}\cdots\ket{\vl^{N-1}}.}
\eeq
Unlike the general state (Eq.\,\ref{eq: combined-ang-mom-states}), this is \resub{involves} only $(N-4)$ intermediate couplings $\vl^{12\ldots}$, since the final CG coefficient in \eqref{eq: Coupling-C} fixes $\ell^{12\ldots(N-2)}_1 = -\ell_1^{N-1}$ and $\ell^{12\ldots(N-2)}_K = \ell_K^{N-1}$ for $K>1$. In total, \resub{this requires $(2N-5)$ indices to fully specify}, regardless of dimension.

\subsection{$(N-1)$-Particle Basis Properties}\label{subsec: multi-basis-properties} The $(N-1)$-particle basis functions of Eqs.\,\ref{eq: combined-ang-mom-states}\,\&\,\ref{eq: zero-ang-mom-states} have analogous properties to those of the single-particle basis (cf.\,\S\ref{sec: single-basis}\ref{subsec: ang-mom-single}). Using \eqref{eq: CG-orthonorm}, we can show orthonormality:
\beq\label{eq: N-1-particle-orthonormality}
    \braket{\resub{\L';\vL'}}{\L;\vL} = \delta^{\rm K}_{\L\L'}\delta^{\rm K}_{\vL\vL'}
\eeq
requiring equality of \resub{both} the combined angular momentum vectors ($\vL$ and $\vL'$) and all components of $\L$ and $\L'$. 
Since the \resub{angular momentum} states form a complete basis, any $(N-1)$-particle function $h:\left(\mathbb{S}^{D-1}\right)^{\otimes (N-1)}$ \resub{can be decomposed into a sum of basis states:}
\beq\label{eq: general-basis-decomp}
    \ket{h} = \sum_{\vL}\sum_{\L}\braket{\resub{\L;\vL}}{h}\ket{\L;\vL} \quad \leftrightarrow \quad h(\hr^1,\ldots,\hr^{N-1}) = \sum_{\vL}\sum_{\L} h_\L^{\vL}\P_\L^{\vL}(\hr^1,\ldots,\hr^{N-1})
\eeq
summing over both combined angular momentum indices $\vL$ and the indices contained within $\L$ (analogously to Eq.\,\ref{eq: 1d-basis-decomp}). The basis components are denoted by $\braket{\resub{\L;\vL}}{h}\equiv h_{\L}^{\vL}$. For the second equality, we switch to wavefunction notation, with the basis functions defined as (cf.\,Eq.\,\ref{eq: combined-ang-mom-states})
\beq\label{eq: N-1-particle-basis-wavefunction}
    \boxed{\P_\L^{\vL}(\hr^1,\ldots,\hr^{N-1}) = \sum_{[\vl^1][\vl^2]\ldots[\vl^{N-1}]}C^{\L;\vL}_{\vl^1\vl^2\ldots\vl^{N-1}}Y_{\vl^1}(\hr^1)Y_{\vl^2}\resub{(\hr^2)}\cdots Y_{\vl^{N-1}}(\hr^{N-1}),}
\eeq
where $Y_{\vl}(\hr)$ are the hyperspherical harmonics of \S\ref{sec: single-basis}\ref{subsec: hypersphericals}. Since $\mathbb{S}^{D-1}$ is the angular part of $\mathbb{M}^D$, \textbf{the directional dependence of any $(N-1)$-particle function in $\mathbb{M}^D$ can be expanded in the separable form of Eq.\,\ref{eq: general-basis-decomp}.}

If the function $h$ appearing in \eqref{eq: general-basis-decomp} has rotational symmetry, a simpler decomposition is possible. In particular, we assume it to be invariant under rotations drawn from \resub{the subspace $\{x_1,\ldots,x_K,0,\ldots,0\}$ of $SO(D)$.} 
\resub{An example of this would be azimuthal symmetry in $D=3$; here, the system is invariant under rotations about one axis only. To obey rotational symmetry in $K$ dimensions, the basis functions must satisfy}
\beq
    \widehat{\mathcal L}^2_K \ket{\L;\vL} = 0 \quad \Rightarrow \quad L_i=0, \quad  \forall \,i < K,
\eeq
where $\widehat{\mathcal L}^2_{K}$ is the combined angular momentum operator (Eq.\,\ref{eq: k-dim-ang-mom}).\footnote{This occurs since any basis function with $L_K\neq 0$ has non-zero combined angular momentum in the $K$-dimensional subspace, violating rotational invariance.} In this instance, only basis functions of the form $\ket{\L;0\ldots 0\; L_K\ldots L_{D-1}}$ enter into \eqref{eq: general-basis-decomp}, \resub{reducing the number of basis coefficients from approximately $\left(\ell_{D-1}^{\rm max}\right)^{2N+D-5}$ to $\left(\ell_{D-1}^{\rm max}\right)^{2N+D-K-5}$, for maximum multipole $\ell_{D-1}^{\rm max}$}. If $h$ is invariant under spatial rotations about \textit{any} axis (\textit{i.e.} it is isotropic), \resub{only the $\vL = \vec 0$ state is required}. In this case
\beq\label{eq: iso-basis-expansion}
    \ket{h} &=& \sum_{\L} \braket{\L;\vec0}{h} \ket{\L;\vec 0} \quad \leftrightarrow \quad h(\hr^1,\ldots,\hr^{N-1}) = \sum_{\L}h_{\L}^{\vec 0} \P_\L^{\vec 0}(\hr^1,\ldots,\hr^{N-1}),
\eeq
for components $h_{\L}^{\vec 0}\equiv \braket{\resub{\L;\vec 0}}{h}$, which may be determined via orthonormality. \textbf{The directional dependence of any \resub{isotropic} $(N-1)$-particle function in $\mathbb{M}^D$ can be expanded in the separable form of Eq.\,\ref{eq: iso-basis-expansion}.}

\resub{For $N=3$ and $N=4$, the isotropic basis functions take the explicit forms:}
\beq
    \P_{\ell_{D-1}}^{\vec 0}(\hr^1,\hr^2) &=& \sum_{\ell_1\ldots\ell_{D-2}} Y_{\ell_1\ell_2\ldots\ell_{D-1}}(\hr^1)Y_{(-\ell_1)\ell_2\ldots\ell_{D-1}}(\hr^2),\\\nonumber
    \P_{\ell^1_{D-1}\ell^2_{D-1}\ell^3_{D-1}}^{\vec 0}(\hr^1,\hr^2,\hr^3) &=& \sum_{\ell_1^1\ldots\ell^1_{D-2}}\sum_{\ell^2_1\ldots\ell^2_{D-2}}\sum_{\ell_1^3\ldots\ell^3_{D-2}}\cg{\vl^1}{\vl^2}{(-\ell^3_1)\ell_2^3\ldots\ell_{D-1}^3}Y_{\ell_1^1\ldots\ell^1_{D-1}}(\hr^1)Y_{\ell_1^2\ldots\ell^2_{D-1}}(\hr^2)Y_{\ell_1^3\ldots\ell^3_{D-1}}(\hr^3).
\eeq
\resub{noting that the final CG coefficient is of the form $\cg{\vl^{12\ldots(N-2)}}{\vl^{N-1}}{\vec 0}$, which enforces $\ell^{12\ldots(N-2)}_1+\ell^{N-1}_1=0$ and $\ell^{12\ldots(N-2)}_K=\ell^{N-1}_K$ for $K>1$. This is a natural extension of the Legendre polynomials to $D$ dimensions; indeed the $D=3$ case recovers the Legendre polynomial $\mathcal{L}_{\ell_2}(\hr^1\cdot\hr^2)$ rescaled by $(-1)^{\ell_2}(4\pi)/\sqrt{2\ell_2+1}$ \citep[cf.][]{2020arXiv201014418C}.}

\resub{Finally, we note} the properties of the basis functions under complex conjugation and parity inversion:
\beq\label{eq: basis-parity-conj}
    \left[\P_\L^{\vec L}(\hr^1,\ldots,\hr^{{N-1}})\right]^* &=& (-1)^{L_{D-1}-L_{1}}(-1)^{\ell^1_{D-1}+\cdots+\ell^{N-1}_{D-1}}\P_\L^{\vec 0}(\hr^1,\ldots,\hr^{{N-1}}) \\\nonumber
    \mathbb{P}\left[\P_\L^{\vec L}(\hr^1,\ldots,\hr^{{N-1}})\right] &=& (-1)^{\ell_{D-1}^1+\cdots+\ell_{D-1}^{{N-1}}}\P_\L^{\vec 0}(\hr^1,\ldots,\hr^{{N-1}}),
\eeq
using \eqref{eq: complex-conj-hypersph}, noting that the CG coefficients enforce $\ell_1^1+\ell_1^2+\cdots+\ell_1^{N-1}+L_1=0$. For $\vL=\vec 0$, this implies that even-(odd-)parity basis functions are purely real (imaginary). 

\section{An Efficient Correlation Function Estimator}\label{sec: NPCF-estimator}
Armed with the $(N-1)$-particle angular basis of \S\ref{sec: npoint-basis}, we now proceed to construct an \resub{efficient} estimator for the $N$-point correlation function \resub{components}. For full generality, we do not assume the NPCF to have any rotational symmetry; such symmetries set various basis components to zero, as discussed in \S\ref{sec: npoint-basis}\ref{subsec: multi-basis-properties}.

\subsection{Derivation}
Assuming statistical homogeneity, the NPCF defined in \eqref{eq: zeta-def} is a function of $(N-1)$ points on $\mathbb{M}^D$, and may thus be expanded in the combined angular momentum basis of \S\ref{sec: npoint-basis} (cf.\,Eq.\,\ref{eq: zeta-basis-decomp-intro}): 
\beq\label{eq: zeta-basis-decomp}
    \boxed{\zeta(\vr^1,\ldots,\vr^{N-1}) = \sum_{\vL}\sum_{\L} \zeta^{\vL}_\L(r^1,\ldots,r^{N-1})\P_{\L}^{\vec L}(\hr^1,\ldots,\hr^{N-1}),}
\eeq
where the basis states, $\P_\L^{\vL}$, are defined in \eqref{eq: N-1-particle-basis-wavefunction}. As before, we sum both over $\vL$, which specifies the properties of the NPCF under joint rotations of all $(N-1)$ direction vectors (with only $\vL = \vec 0$ required if the NPCF is isotropic), and $\L \equiv \{\ell^1_{D-1},\ell_{D-1}^2,\ell_{D-1}^{12},\ell^3_{D-1},\resub{\ell^{123}_{D-1},\ell^4_{D-1}},\ldots,\ell_{D-1}^{{N-1}}\}$, which defines the relative orientations of the direction vectors. In this form, the NPCF is fully specified by the \resub{basis coefficients} $\zeta^{\vL}_\L$, which are functions only of the radial parameters $r^i$. 

Due to the parity properties of the basis functions (Eq.\,\ref{eq: basis-parity-conj}), basis coefficients with even (odd) $\sum_{i=1}^{N-1} \ell_{D-1}^i$ represent even-parity (odd-parity) NPCF contributions; furthermore, they are purely real (imaginary) if the random field $X$ is real-valued and $\vL = \vec 0$. Parity-odd isotropic basis functions occur only for $N>D$; at lower orders, a parity transformation is equivalent to a rotation, under which the basis functions are invariant.

The basis coefficients can be extracted from \eqref{eq: zeta-basis-decomp} via an inner product:
\beq\label{eq: zeta-basis-coeff}
    \zeta_\L^{\vL}(r^1,\ldots,r^{N-1}) \equiv \braket{\L;\vL}{\zeta} =  \int_{\left(\mathbb{S}^{D-1}\right)^{\otimes (N-1)}}d\Omega_{D-1}^1\cdots d\Omega_{D-1}^{N-1}\,\left[\zeta(\vr^1,\ldots,\vr^{N-1})\P_\L^{\vec L,*}(\hr^1,\ldots,\hr^{N-1})\right],
\eeq
where the integral is over $(N-1)$ copies of the angular space. As in \eqref{eq: NPCF-estimator-full}, the NPCF may be estimated as a product of $N$ random fields, integrated over space; inserted into \eqref{eq: zeta-basis-coeff} this yields
\beq\label{eq: NPCF-estimator-basis-1}
    \hat\zeta^{\vL}_\L(r^1,\ldots,r^{N-1}) = \frac{1}{V_D}\int_{\mathbb{M}^D} d^D\vs \int_{\left(\mathbb{S}^{D-1}\right)^{\otimes (N-1)}}d\Omega_{D-1}^1\cdots d\Omega_{D-1}^{N-1}\,\left[X(\vs)X(\vs+\vr^1)\cdots X(\vs+\vr^{N-1})\P_\L^{\vec L,*}(\hr^1,\ldots,\hr^{N-1})\right].
\eeq
Finally, we insert the explicit forms of the $(N-1)$-particle basis functions (Eq.\,\ref{eq: N-1-particle-basis-wavefunction}), which \resub{yields}
\beq\label{eq: NPCF-mult-estimator}
    \boxed{\hat\zeta^{\resub{\vec L}}_\L(r^1,\ldots,r^{N-1}) = \frac{1}{V_D}\sum_{[\vl^1]\ldots[\vl^{N-1}]}C^{\L;\vL}_{\vl^1\ldots\vl^{N-1}}\int_{\mathbb{M}^D} d^D\vs\,\left[X(\vs)a_{\vl^1}(\vs;r^1)\cdots a_{\vl^{N-1}}(\vs;r^{N-1})\right],}
\eeq
defining the functions
\beq\label{eq: alm-def}
    a_{\vl}(\vs;r) \equiv \int_{\mathbb{S}^{D-1}}d\Omega_{D-1}\,X(\vs+\vr)Y^*_{\vl}(\hr).
\eeq
\resub{Importantly, the angular integrals are now fully decoupled.}


\resub{Usually, the NPCF statistic is binned in radius via a set of $(N-1)$ top-hat filters, $\Theta^{b^K}(r^K)$, which are equal to $1$ if $r^K$ is in bin $b^K$ and zero else. In this case, $\hat\zeta_{\L}^{\vL}(r^1,\ldots,r^{N-1})$ is replaced by its binned form $\hat\zeta_{\L}^{\vL,\vec b}$, where $\vec b \equiv\{b^1,\ldots,b^{N-1}\}$. This is again estimated using \eqref{eq: NPCF-mult-estimator}, defining the bin integrated functions
\beq\label{eq: alm-bin-def}
    a_{\vl}^{b}(\vec s) = \frac{1}{v_b}\int_{\mathbb{M}^D}d^D\vr\,X(\vs+\vr)Y_{\vl}^*(\hr)\Theta^b(r),
\eeq
where $v_b \equiv \int_{\mathbb{M}^D}d^D\vr$ is the bin volume. Assuming a fixed maximum multipole $\ell_{D-1}^{\rm max}$ and some number of bins $N_b$, this ensures that only a finite number of $a_{\vl}^{b}(\vs)$ coefficients (asymptotically, $N_b\times\left(\ell_{D-1}^{\rm max}\right)^{N-1}$) need to be estimated at each position $\vs$. If one wished to reconstruct the full correlator $\zeta(\vr^1,\ldots,\vr^{N-1})$ from the set of measured basis coefficients, this truncation would lead to an approximation error. In practice, this can be avoided by projecting the theory model in the same manner as the data; then, using a low $\ell_{D-1}^{\rm max}$ will lead only to a slight loss of information, depending on the model in question.}

Estimation of the NPCF basis components reduces to two operations: (1) computing \resub{$a_{\vl}^b(\vs)$ for each radial bin $b$} and angular momentum eigenvalue set $\vl$ of interest, 
and (2) performing a spatial integral over $\vs$, alongside a sum over the lower-dimensional angular momentum eigenvalues \resub{$[\vl^i]\equiv\{\ell^i_1,\ldots,\ell^i_{D-2}\}$}, subject to the coupling rules of \eqref{eq: selection-rules}\,\&\,\eqref{eq: addition-triangle-conditions}. Computationally, this is \resub{much} more efficient than a direct implementation of \eqref{eq: NPCF-estimator-basis-1}. A cartoon indicating this procedure for $N=4$ is shown in Fig.\,\ref{fig: npcfs}.

\subsection{Application to Discrete Data}\label{subsec: discrete-estimator}
\resub{For discrete data, the random field $X$ can be represented as a (weighted) sum of Dirac deltas, as in \eqref{eq: NPCF-estimator-intro-discrete}. Inserting this definition} into \eqref{eq: NPCF-mult-estimator} leads to the following estimator \resub{for the NPCF basis coefficients in bins $\vec b$}:
\beq\label{eq: NPCF-estimator-discrete}
    \resub{\hat\zeta_\L^{\vL,\vec b}} &=& \frac{1}{V_D}\sum_{j=1}^n \resub{w^j}\sum_{[\vl^1]\ldots[\vl^{N-1}]}C^{\L;\vL}_{\vl^1\ldots\vl^{N-1}}\left[\resub{a^{b^1}_{\vl^1}(\vy^j)\cdots a_{\vl^{N-1}}^{b^{N-1}}(\vy^j)}\right],\quad    a_{\vl}^{b}(\vy^j) = \frac{1}{v_b}\sum_{k=1}^n\,w^kY^*_{\vl}(\widehat{\vy^k-\vy^j})\Theta^{b}\left(\left|\vy^k-\vy^j\right|\right).
\eeq
Practically, the \resub{$a_{\vl}^b(\vy^j)$} functions may be computed by summing over $n$ points, $\{\vy^k\}$, weighted by a hyperspherical harmonic and a \resub{binning function} in the separation vector $\vy^k-\vy^j$. Since the functions must be estimated at the location of each primary particle, $\vy^j$, \textbf{the algorithm has complexity $\mathcal{O}(n^2)$} \resub{(with respect to $n$), for any choice of $D$ or $N$}. This is \resub{significantly} faster than the na\"ive NPCF estimator of \eqref{eq: NPCF-estimator-intro-discrete}, \resub{inserting the basis functions}
\beq\label{eq: B-hypersph}
    \resub{B(\vr^1,\ldots,\vr^{N-1})\equiv \P_{\vec\Lambda}^{\vL}(\hr^1,\ldots,\hr^{N-1})\Theta^{b^1}(r^1)\cdots\Theta^{b^{N-1}}(r^{N-1}),}
\eeq
\resub{showing the utility of our hyperspherical harmonic decomposition.}

\resub{Although the scaling with $n$ is independent of $D$ and $N$, we do expect the computational cost of a full NPCF measurement to increase in higher-dimensional scenarios. In part, this occurs since the number of intermediate $[\vl^i]$ summations in \eqref{eq: NPCF-estimator-discrete} is a strong function of $D$ and $N$. In total, we must sum over approximately $(D-1)(N-1)$ indices, each of which is bounded by $\ell^{\rm max}_{D-1}$, thus the computation time is asymptotically $\propto\left(\ell^{\rm max}_{D-1}\right)^{(D-1)(N-1)}$. Since this summation must be done once per primary particle $\vy^j$, we expect the algorithm to scale linearly with $n$ if this process dominates over $a_{\vl}^b$ computation. Secondly, the number of basis vectors at fixed $\ell^{\rm max}_{D-1}$ is exponential in $D$ and $N$. Asymptotically, this scales as $N_b^{N-1}\times \left(\ell_{D-1}^{\rm max}\right)^{2N-4}\times \left(\ell_{D-1}^{\rm max}\right)^{D-1}$ (counting elements of $\vec b, \vec \Lambda$ and $\vec L$ respectively, using $N_b$ radial bins per dimension). Although this is a strong scaling, it is generic to any higher-point basis (and usually referred to as the `curse of dimensionality'). In the presence of certain symmetries, the number of basis functions is significantly reduced: for example, isotropy demands that $\vL=\vec 0$, reducing the scaling to $N_b^{N-1}\times \left(\ell_{D-1}^{\rm max}\right)^{2N-5}$, independent of $D$. In the below, we will always compare na\"ive and efficient estimators projected onto the same basis, thus this factor appear in the ratio of computation times.}

\subsection{Application to Gridded Data}\label{subsec: gridded-estimator}
Our \resub{estimators may} also be applied to continuous data discretized on some regular grid, which is of use for the analysis of hydrodynamic simulations, for example. For this, we first rewrite \resub{$a_{\vl}^b(\vs;r)$} (Eq.\,\ref{eq: alm-bin-def}) as
\beq
    \resub{a_{\vl}^b(\vs) = \int_{\mathbb{M}^{D}}d^D\vr\,X(\vs+\vr)Y^*_{\vl}(\hr)\Theta^b(r) \equiv (-1)^{\ell_{D-1}}\int_{\mathbb{M}^{D}}d^D\vr\,X(\vs-\vr)\left[Y_{\vl}(\hr)\Theta^b(r)\right],}
\eeq
\resub{relabelling variables and utilizing the conjugate properties of hyperspherical harmonics (Eq.\,\ref{eq: complex-conj-hypersph}).}
For gridded data in Euclidean space, \textit{i.e.} with $\mathbb{M}^D = \mathbb{R}^{D}$, this may be straightforwardly computed using the $D$-dimensional FFT. Explicitly:
\beq
    \resub{a_{\vl}^b(\vs) = (-1)^{\ell_{D-1}}\mathrm{FFT}^{-1}\left[\mathrm{FFT}(X)\mathrm{FFT}(Y_{\vl}\;\Theta^b)\right].}
\eeq
where $\mathrm{FFT}^{-1}$ is the inverse FFT. These operations have complexity $\mathcal{O}(n_{\rm g}\log n_{\rm g})$ for $n_{\rm g}$ grid points. Following computation of the various $a_{\vl}^{b}(\vs)$ terms, the \resub{estimator for the NPCF components} can be constructed from \eqref{eq: NPCF-mult-estimator} as a simple sum in $D$ dimensions. \textbf{The full estimator has complexity $\mathcal{O}(n_{\rm g}\log n_{\rm g})$}, which is again much faster than the na\"ive $\mathcal{O}(n_{\rm g}^N)$ \resub{result. The above procedure may also be applied to the discrete data-sets discussed in \S\ref{subsec: discrete-estimator}, via a non-uniform FFT \citep{nuFFT}.}


\section{Applications}\label{sec: applications}
We now consider a number of physical scenarios in which the above methods can be employed, \resub{and give numerical examples. For this purpose, we provide a \textsc{Julia} implementation of the two main algorithms discussed above (\ref{eq: NPCF-estimator-intro-discrete}\,\&\,\ref{eq: NPCF-estimator-discrete}).\footnote{\href{https:/github.com/oliverphilcox/NPCFs.jl}{GitHub.com/oliverphilcox/NPCFs.jl}} This can compute the NPCF of discrete particles with $N\in\{2,3,4,5\}$, using Cartesian geometries with $D \in\{2,3,4\}$ or spherical geometries with $D=2$ and is fully parallelized.}

\begin{figure}[t]
    \subfloat[4PCF in $\mathbb{S}^2$]{%
      \includegraphics[width=0.47\textwidth]{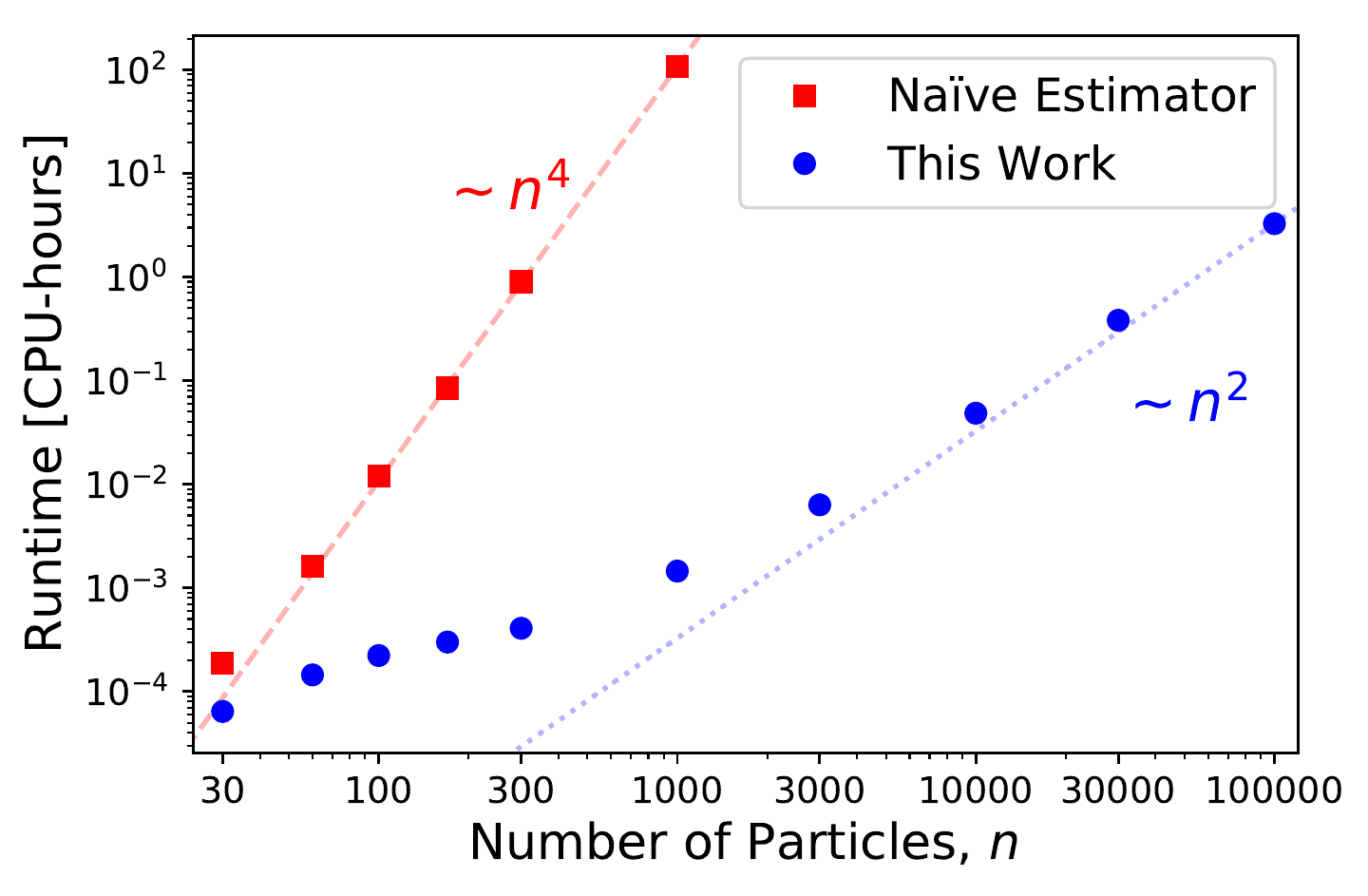}
    }
    \hfill
    \subfloat[5PCF in $\mathbb{R}^3$]{%
       \includegraphics[width=0.46\textwidth]{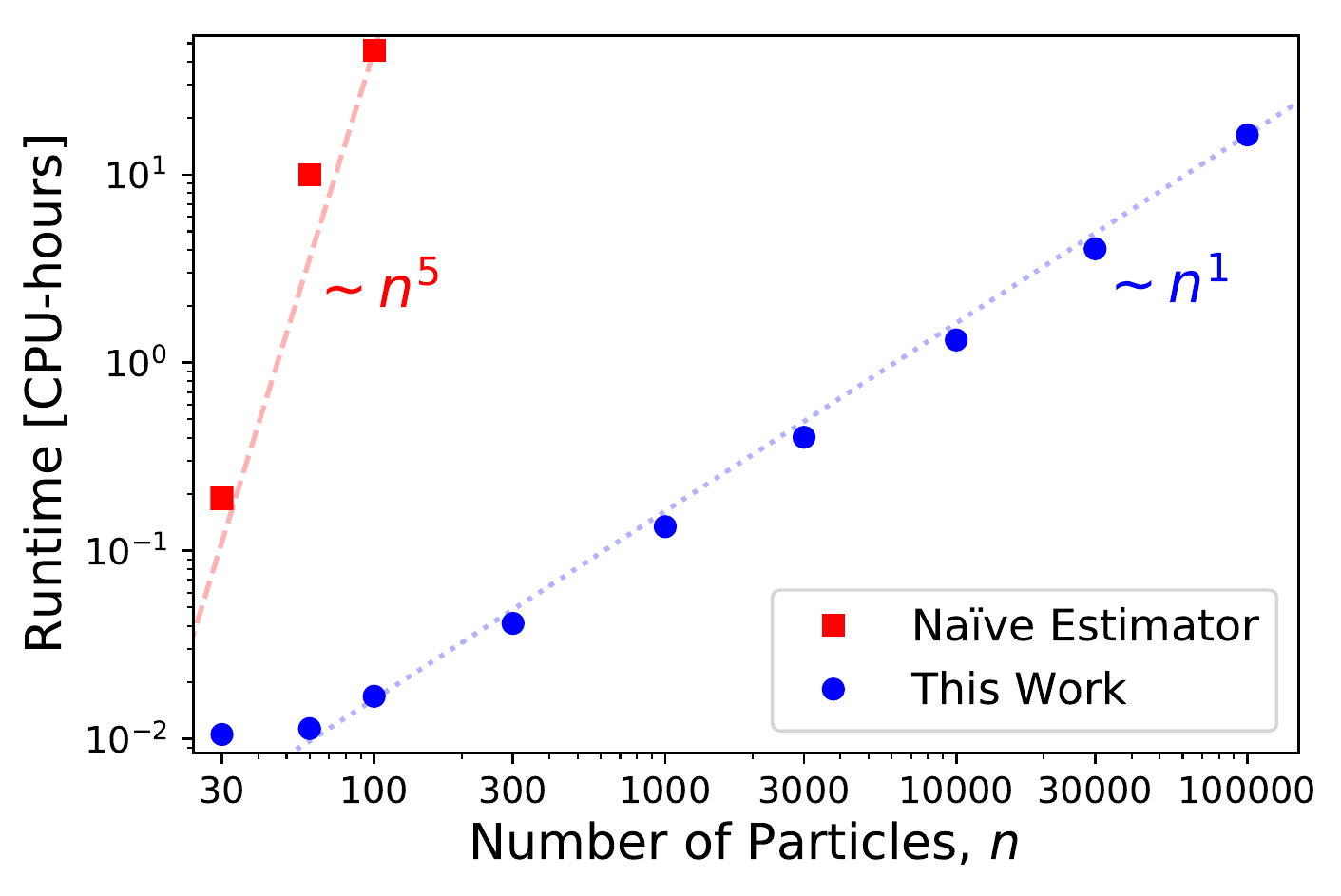}
    }
    \caption{\resub{Timing comparison of the na\"ive NPCF estimator with that introduced in this work. In the first case, the NPCF is estimated by counting sets of $N$ particles, via \eqref{eq: NPCF-estimator-intro-discrete}, weighting by the relevant basis functions (\S\ref{sec: npoint-basis}). The new estimators exploit hyperspherical harmonic decompositions to reduce the estimator to a sum over pairs. Here, we show results for a variety of choices of $n$, for both 4PCF estimation on the two-sphere (left), and 5PCF estimation in 3D Cartesian space (right). As expected, the runtime of na\"ive estimator scales as $n^N$ (as indicated by the red dotted lines), but the new estimators scale as $n^2$ for the 4PCF, or $n$ for the 5PCF. The latter scaling arises since we are dominated by the sum over intermediate momenta $[\vec\ell^i]$, rather than the sum over pairs; at larger $n$, we expect a quadratic scaling with $n$. All computations were performed in \textsc{Julia} using 16 CPUs, and we have verified that the measured NPCF components agree to machine precision}.}
    \label{fig: npcf-timings}
\end{figure}

\subsection{Cosmic Microwave Background}
Cosmic Microwave Background (CMB) radiation encodes a snapshot of the Universe at the epoch of recombination, around $380,000$ years after the Big Bang. This may be probed using microwave satellites such as \textit{WMAP} and \textit{Planck}, which map the CMB temperature fluctuations as a function of direction; such observations have been used to place strong constraints on cosmological parameters such as the matter density and Universe's expansion rate \citep[e.g.,][]{2011ApJS..192...18K,2020A&A...641A...6P}. 

In this \resub{setting}, the random field in question is the fractional temperature fluctuation on the 2-sphere, $\Theta:\mathbb{S}^2\to\mathbb{R}$, \textit{i.e.} $\mathbb{M}^D = \mathbb{S}^2$. Conventionally, $\Theta$ is expanded in $D=3$ spherical harmonics as \resub{$\Theta(\phi,\theta;\hn) = \sum_{\lambda=0}^\infty\sum_{\mu=-\lambda}^{\lambda}\Theta_{\lambda\mu}Y_{\lambda\mu}(\phi,\theta)$}, using spherical polar coordinates $\phi\in[0,2\pi)$, $\theta\in[0,\pi)$ relative to some pole on the sphere at position vector $\hn$. The statistical properties of $\Theta$ are then characterized in terms of the spherical harmonic coefficients \resub{$\Theta_{\lambda\mu}$ (often denoted $a_{\lambda\mu}^T$)}. To apply the techniques of this work, we instead expand the temperature fluctuations using the $D=2$ hyperspherical harmonics (\S\ref{sec: single-basis}\ref{subsec: hypersphericals}), \textit{i.e.}
\beq
    \Theta(\vr;\hn) = \sum_{\ell=0}^\infty \Theta_\ell(\theta)Y_{\ell}(\phi) \equiv \frac{1}{\sqrt{2\pi}}\sum_{\ell=0}^\infty \Theta_\ell(\theta)e^{i\ell\phi},
\eeq
identifying $\ell$ as the total angular momentum and $\theta$ as the radial coordinate relative to $\hn$, which acts as an origin on $\mathbb{S}^2$.\footnote{Here, $\theta$ measures the arc-lengths of great-circles through two points on the 2-sphere, as viewed in $\mathbb{R}^3$.} This is a convenient basis for computing higher-order clustering statistics on the two-sphere, since (a) it avoids the need for an embedding space, and (b), it provides a natural split into isotropic and anisotropic correlators.

As in \eqref{eq: zeta-def}, the temperature NPCF is defined as a statistical average over $\Theta$:
\beq\label{eq: cmb-NPCF-def}
    \zeta(\vr^1,\ldots,\vr^{N-1}) = \mathbb{E}_\Theta\left[\Theta(\bvec 0;\hn)\Theta(\vr^1;\hn)\cdots \Theta(\vr^{N-1};\hn)\right];
\eeq
by statistical homogeneity, this is independent of the choice of origin $\hn$. \eqref{eq: cmb-NPCF-def} may be expanded \resub{in the basis of \eqref{eq: general-basis-decomp}}, where the $D=2$ basis functions take the form
\beq
    \resub{\P^{L}_{\ell^1\ldots\ell^{N-1}}}(\phi^1,\ldots,\phi^{N-1}) &=& (2\pi)^{-(N-1)/2}\mathrm{exp}\left[i(\ell^1\phi^1+\cdots+\ell^{N-1}\phi^{N-1})\right],
\eeq
\resub{with $\ell^1+\cdots+\ell^{N-1} = L$, where $L=0$ for isotropic correlators}. Note that no intermediate angular momenta need to be specified due to the coupling rules of \eqref{eq: addition-triangle-conditions}. As in \eqref{eq: NPCF-mult-estimator}, we may form an $\mathcal{O}(n^2)$ estimator \resub{for the NPCF coefficients}:
\beq
    \zeta_{\ell^1\ldots\ell^{N-1}}^{L}(\theta^1,\ldots,\theta^{N-1}) &=& \frac{1}{4\pi}\int_{\mathbb{S}^2} d\Omega_2\,\left[\Theta(\hn)a_{\ell^1}(\hn;\theta^1)\cdots a_{\ell^{N-1}}(\hn;\theta^{N-1})\right],\\\nonumber
    a_{\ell}(\hn;\theta) &\equiv& \frac{1}{\sqrt{2\pi}}\int_0^{2\pi} d\phi\, \Theta(\phi,\theta;\hn)e^{i\ell\phi}.
\eeq
The first integral is over all possible choices of origin $\hn$, whilst the second is over a circle centered at $\hn$ with radial parameter $\theta$ \resub{(which may be discretized into bins, as before)}.

Such estimators are straightforward to implement and allow efficient computation of the higher-order CMB NPCFs, albeit in a basis somewhat different to that usually adopted. We caution that the $\ell$ indices play a different \resub{role} to those of the \resub{$\lambda$} indices appearing in the standard spherical harmonic expansion of $\Theta$. In our basis, $\ell$ represents angular momentum around an origin \textit{on} the 2-sphere, whilst \resub{$\lambda$} is with reference to the origin \textit{of} the 2-sphere in the embedding space $\mathbb{R}^3$. In practice, this allows us to restrict to \resub{much} smaller $\ell$ than used conventionally.\footnote{CMB fluctuations have characteristic angular scale $r_s(z_*)/d_A(z_*)$ where $r_s$ is the sound horizon in comoving coordinates, $d_A$ is the angular diameter distance and $z_*$ is the redshift at the end of the baryon drag epoch. This imprints a characteristic angular momentum scale $L\sim \pi d_A(z_*)/r_s(z_*)\gg 1$. In our basis, $\ell$ corresponds to the ratio of two polygon sides on $\mathbb{S}^2$ and is thus $\mathcal{O}(1)$.} We further note that the CMB contains also \resub{polarization} fluctuations. These may be analyzed using an extension of the above formalism, replacing the hyperspherical harmonics with spin-weighted hyperspherical harmonics \citep[e.g.,][]{2012PhRvD..86l5013D,2016JMP....57i2504B}.

\resub{To give a sense of how the above algorithms work in CMB contexts, we consider a simple problem: estimating the isotropic 4PCF of randomly placed points on the two-sphere. This corresponds to the above scenarios with $N=4$, $D=2$ and a spherical geometry. For this test, we generate a set of $n$ evenly distributed points, and compute the coefficients $\zeta_{\ell^1\ell^2\ell^3}^{0,b^1b^2b^3}$, using $10$ radial bins per dimension with $\cos\theta\in[-0.5,0.5)$ and $\ell_{\rm max}=4$. This leads to a total of $35$ ($120$) angular (radial) components. Timing results for the computation using both the na\"ive and quadratic estimators (projecting the 4PCF onto the same basis functions in both cases) are shown in Fig.\,\ref{fig: npcf-timings}(a) for a variety of choices of $n$, using our public \textsc{Julia} code. As expected, the runtime scales as $n^4$ for the na\"ive estimator, which leads to unwieldy computation times even for a few thousand particles. For the estimators introduced in this work, the runtime scales as $n^2$ for large $n$, as expected from the algorithm's complexity.}

\subsection{Hydrodynamic Turbulence}\label{subsec: hydro-turb}
NPCFs have found significant use in the study of hydrodynamical turbulence. Being a chaotic process, the evolution of the velocity and density fields in a turbulent flow cannot be treated deterministically; rather, they must be analyzed statistically. Furthermore, the density fields of turbulent media are known to be close to log-normal \citep{1994ApJ...423..681V}, implying that the higher-order NPCFs functions contain a wealth of information, particularly concerning the sonic and Alfv\'enic Mach numbers \citep[e.g.,][]{2007ApJ...658..423K,2009ApJ...693..250B,2010ApJ...708.1204B,2018ApJ...862..119P}. 

\resub{One of the simplest observables} is the turbulent density field, $\rho:\mathbb{R}^3\to\mathbb{R}$, whose $N$-point function is defined in \eqref{eq: zeta-def}. In the absence of any external forcing, we expect the NPCF to be statistically isotropic, thus it can be expanded via \eqref{eq: iso-basis-expansion} in terms of the $D=3$ basis functions \resub{with $\vL=\vec 0$. As shown in \S\ref{sec: single-basis}\ref{subsec: hypersphericals},} the corresponding one-particle basis functions are just the usual spherical harmonics, $Y_{m\ell}(\hr)$, and their coupling can be expressed in terms of Wigner 3-$j$ symbols. 

As before, we may form $\mathcal{O}(n^2)$ estimators \resub{for the NPCF coefficients} via \eqref{eq: NPCF-mult-estimator}. \resub{As an example, the isotropic 5PCF estimator becomes (ignoring radial binning for clarity)}
\beq\label{eq: D=3-NPCF}
    \zeta^{\vec 0}_{\ell^1\ell^2\ell^{12}\ell^3\ell^4}(r^1,r^2,r^3) &=& \frac{1}{V_3}(-1)^{\ell^1+\ell^2+\ell^3+\ell^4}\sum_{m^1m^2m^3m^4} (-1)^{\ell^{12}-m^{12}}\sqrt{2\ell^{12}-1}\tj{\ell^1}{\ell^2}{\ell^{12}}{m^1}{m^2}{-m^{12}}\tj{\ell^{12}}{\ell^3}{\ell^4}{m^{12}}{m^3}{m^4}\nonumber\\
    &&\,\times\,\int_{\mathbb{R}^3} d\vs\,\left[\rho(\vs)a_{m^1\ell^1}(\vs;r^1)a_{m^2\ell^2}(\vs;r^2)a_{m^3\ell^3}(\vs;r^3)a_{m^4\ell^4}(\vs;r^4)\right],
\eeq
where $V_3$ is the volume of the space, \resub{$m^{12}\equiv m^1+m^2$, and $a_{m\ell}(\vs; r) = \int_{\mathbb{S}^2}d\Omega_2\,\rho(\vs+\vr)Y^*_{m\ell}(\hr)$.} 
Introducing spin-weighted (or vector) spherical harmonics, the approach may be extended to tensorial correlators, such as those of the velocity field.

\resub{Fig.\,\ref{fig: npcf-timings}(b) presents a practical demonstration of the isotropic 5PCF estimator, applied to $n$ discrete points in 3D. We consider ten radial bins in $[0.1,0.4]$ for uniformly distributed data in a periodic cube of length $1$, and fix $\ell_{\rm max}=4$. In this case, the 5PCF is specified by four radial bin indices and five angular multiplets, as in \eqref{eq: D=3-NPCF}. This gives a total of $210$ radial and $585$ angular components. As before, we find that the runtime of the na\"ive estimator scales as $n^N$, which quickly becomes computationally prohibitive. For the NPCF estimator of this work, the runtime appears to be linear in $n$, rather than quadratic: this occurs since the work is dominated by the $m^i$ summations (cf.\,\S\ref{sec: NPCF-estimator}\ref{subsec: discrete-estimator}), though we expect the $n^2$ scaling to dominate for denser samples. In all cases however, our approach is significantly faster than that of the na\"ive estimator. We note that our algorithm can be further accelerated by gridding the data, and making use of Fourier transforms.}

\subsection{Large-Scale Structure}
The large-scale distribution of matter in the late Universe follows a weakly non-Gaussian distribution and is commonly analyzed using $N$-point statistics \citep[e.g.,][]{2001ASPC..252..201P}. The underlying space is expected to be flat, homogeneous, and isotropic, and is thus described by the metric of \eqref{eq: metric} with $k=0$ and $D=3$.\footnote{Our methodology applies similarly to $k\neq 0$, though there is significant evidence implying that the Universe is flat \citep{2020A&A...641A...6P}. Additionally, our methods can be used to compute \resub{projected} correlation functions in $\mathbb{R}^2$, requiring the $D=2$ basis functions, as demonstrated in \citep{2015MNRAS.454.4142S} for the 3PCF.} A common task in cosmology is the estimation of isotropic NPCFs for the galaxy overdensity field, $\delta_g:\mathbb{R}^3\to\mathbb{R}$; this proceeds identically to \S\ref{sec: applications}\ref{subsec: hydro-turb}, except that the data are discrete. Full discussion of this (including implementation in the \textsc{encore} code\footnote{\href{https://GitHub.com/oliverphilcox/encore}{GitHub.com/oliverphilcox/encore}}) is presented in our companion work \citep{npcf_algo}, and allows information to be extracted from the high-order NPCFs, which are otherwise computationally prohibitive to measure.

Due to the effects of redshift-space distortions \citep[e.g.,][]{1987MNRAS.227....1K}, observed galaxy distributions are \resub{not} isotropic, implying that the decomposition of \eqref{eq: iso-basis-expansion} does not capture all possible NPCF information. However, the statistics are invariant under rotations about an (assumed fixed) line-of-sight, here set to $\hat{\bvec x}^3$. For a full treatment, we must instead expand the NPCF using Eq.\,\ref{eq: general-basis-decomp}, keeping only terms with $L_1=0$ (cf.\,\S\ref{sec: npoint-basis}\ref{subsec: multi-basis-properties}):
\beq
    \zeta(\vr^1,\ldots,\vr^{N-1}) = \sum_{L=0}^\infty \sum_{\L} \zeta_\L^{0L}(r^1,\ldots,r^{N-1})\P^{0L}_{\L}(\hr^1,\ldots,\hr^{N-1}),
\eeq
writing $L \equiv L_2$. As an example, the 3PCF becomes 
\beq
    \zeta(\vr^1,\vr^2) &=& \sum_{L=0}^\infty \sum_{\ell^1\ell^2}\zeta^{0L}_{\ell^1\ell^2}(r^1,r^2)\P^{0L}_{\ell^1\ell^2}(\hr^1,\hr^2), \\\nonumber
    \P^{0L}_{\ell^1\ell^2}(\hr^1,\hr^2) &=& (-1)^{\ell^2-\ell^1}\sqrt{2L+1}\sum_m \tj{\ell^1}{\ell^2}{L}{m}{-m}{0} Y_{\ell^1m}(\hr^1)Y_{\ell^2-m}(\hr^2).
\eeq
\cite[cf.\,][]{2018MNRAS.478.1468S,2019MNRAS.484..364S}. Such statistics may be estimated via \eqref{eq: NPCF-mult-estimator}, as before. The above decomposition provides a complete basis for the redshift-space 3PCF \citep[analogous to][]{2019MNRAS.484..364S} and extends naturally to higher orders, \resub{which have not previously been discussed.}

\section{Summary}

Many areas of research require computation of clustering statistics from continuous or discrete random fields. Perhaps the most prevalent statistic is the $N$-point correlation function (NPCF), defined as the statistical average over $N$ fields in different spatial locations. If the random field is Gaussian-distributed, only the 2PCF is of interest; in the general case, all correlators have non-trivial forms. Given a set of $n$ particles, a na\"ive estimator \resub{for the NPCF components in some basis} has $\mathcal{O}(n^N)$ complexity \resub{with respect to $n$}. As $N$ increases, this rapidly becomes computationally infeasible to apply: alternative methods must be sought if one wishes to unlock the information contained within higher-order NPCFs.

This work considers NPCF estimation on isotropic and homogeneous manifolds in $D$ dimensions. Under these assumptions (which encompass spherical, flat, and hyperbolic geometries), we show that any function of one position can be expanded in hyperspherical harmonics; a $D$-dimensional analog of the conventional spherical harmonics. These are also eigenstates of the angular momentum operators; utilizing the mathematics of angular momentum addition, we can construct basis functions involving $(N-1)$ points on  $\mathbb{S}^{D-1}$ as a sum over products of $(N-1)$ hyperspherical harmonics. This forms a natural angular basis for the NPCF, particularly if the random field is statistically isotropic. The decomposition allows construction of an NPCF estimator that separates into a product of $(N-1)$ spatial integrals; \textbf{this has $\mathcal{O}(n^2)$ \resub{complexity}} \resub{(with respect to $n$)}, or $\mathcal{O}(n_{\rm g}\log n_{\rm g})$ using FFTs with $n_{\rm g}$ grid-points. \resub{The algorithms have been validated numerically using a new \textsc{Julia} implementation; in all scenarios tested, we find our approach to yield significantly faster measurements of the NPCF coefficients in our angular momentum basis.}

Such techniques will allow high-order correlation functions to be computed from data, allowing more complete analysis of phenomena ranging from fluid turbulence to galaxy clustering. \resub{Furthermore, since the algorithm can be applied to scenarios with $D\neq 3$, we may consider also the computation of NPCFs on the surface of spheres (relevant e.g., for atmospheric physics), or in higher-dimensional atomic treatments with $D=6$ \citep[e.g.,][]{cooper1963}.} 
These ideas may be extended further; a case of particular interest is in the correlation functions of random fields with non-zero spin; these are required to describe the statistics of \resub{tensor} fields such as turbulent velocities and CMB polarization.

\acknow{We thank David Spergel for encouraging us to write this paper, as well as Robert Cahn, Simone Ferraro, Jiamin Hou,  \resub{Bhuv Jain, Adam Lidz}, Moritz M\"unchmeyer, \resub{Shivam Pandey}, Ue-Li Pen, \resub{Cristiano Sabiu}, Frederik Simons, Sauro Succi, and Wen Yan for insightful discussions. \resub{We are additionally indebted to the anonymous referees for insightful feedback.} OP acknowledges funding from the WFIRST program through NNG26PJ30C and NNN12AA01C \resub{and thanks the Simons Foundation for additional support}.
}

\showacknow 

\appendix

\section{Hyperspherical Harmonics}\label{appen: hyperspherical}
\resub{The hyperspherical harmonics are defined by the eigenfunction equation \eqref{eq: eigenfunction-S-D-1}, which can be written explicitly as}
\beq\label{eq: eigenfunction-eqns}
    \Delta_{\mathbb{S}^{D-1}}Y(\hr) \equiv \sin^{2-D}\theta_1\,\partial_{\theta_1}\left[\sin^{D-2}\theta_1\,\partial_{\theta_1}Y(\hr)\right]+\cdots +  \sin^{-2}\theta_1\cdots \sin^{-2}\theta_{D-2}\,\partial_{\theta_{D-1}}^2Y(\hr) &=& -\lambda_{D-1} Y(\hr)
\eeq
in hyperspherical coordinates. Importantly, the Laplace-\resub{Beltrami} operator on the $K$-sphere may be written in terms of that on the $(K-1)$-sphere, allowing us to compute the angular basis functions iteratively, first solving for the 1-sphere basis, which can then be used to find the solution on the 2-sphere, \textit{et cetera}. \resub{This leads to the hierarchy $Y^{(K)}(\theta_1,\ldots,\theta_{K}) = \Theta_{K}(\theta_{K})Y^{(K-1)}(\theta_1,\ldots,\theta_{K-1})$, where}
\beq\label{eq: hypersph-eqn}
    &&\sin^{1-K}\theta_{K}\,\frac{\partial}{\partial \theta_{K}}\left[\sin^{K-1}\theta_{K}\,\frac{\partial \Theta_{K}(\theta_{K})}{\partial \theta_{K}}\right]-\lambda_{K-1}\sin^{-2}\theta_{K}\,\Theta_{K}(\theta_{K}) = - \lambda_{K}\Theta_K(\theta_K)
\eeq
\citep{higuchi1987}, \resub{writing} the eigenvalue for the $K$-sphere as $\lambda_{K}$. Starting from the 1-sphere basis functions $\Theta_1(\theta_1) = (2\pi)^{-1/2}\mathrm{exp}\left(-i\ell_1\theta_1\right)$, we can form the full hyperspherical harmonics by solving \eqref{eq: hypersph-eqn}, giving $Y_{\ell_1\ldots\ell_{D-1}}(\theta_1, \ldots,\theta_{D-1}) \equiv \Theta_1(\theta_1)\Theta_2(\theta_2)\cdots\Theta_{D-1}(\theta_{D-1})$, with 
\beq\label{eq: hypersph-def}
    Y_{\ell_1\ldots\ell_{D-1}}(\theta_1, \ldots,\theta_{D-1}) &\equiv& \Theta_1(\theta_1)\Theta_2(\theta_2)\cdots\Theta_{D-1}(\theta_{D-1}),\\\nonumber
    \Theta_K(\theta_K) &=& \sqrt{\frac{2\ell_K+K-1}{2}\frac{(\ell_K+\ell_{K-1}+K-2)!}{(\ell_K-\ell_{K-1})!}}\sin^{1-K/2}\theta_K\,P_{\ell_K+K/2-1}^{-(\ell_{K-1}+K/2-1)}\left(\cos\theta_{K}\right), \quad (2\leq K\leq D-1),
\eeq
where $P_L^M(x)$ is an associated Legendre polynomial \citep[e.g.,][\S14.3]{NIST:DLMF}. \resub{The indices $\ell_i$ are related to the} Laplace-\resub{Beltrami} eigenvalues of \eqref{eq: laplace-bertrami-eq} via $\lambda_K = \ell_K(\ell_K+K-1)$ \resub{and must satisfy the selection rules of \eqref{eq: selection-rules}.}

Under complex conjugation and parity transformations, the hyperspherical harmonics transform as
\beq\label{eq: complex-conj-hypersph}
    Y_{\ell_1\ell_2\ldots\ell_{D-1}}^*(\hr) = (-1)^{\ell_1}Y_{(-\ell_1)\ell_2\ldots\ell_{D-1}}(\hr), \qquad     \mathbb{P}\left[Y_{\ell_1\ldots\ell_{D-1}}(\hr)\right] = (-1)^{\ell_{D-1}}Y_{\ell_1\ldots\ell_{D-1}}(\hr),
\eeq
where the parity operator, $\mathbb{P}$, sends $\vr\to-\vr$ and thus $\theta_1\to\pi+\theta_1$, $\theta_k\to\pi-\theta_k$ for $k>1$. The latter equation can be verified by noting that the action of the parity operator on the basis components of \eqref{eq: hypersph-def} are $\mathbb{P}[\Theta_1(\theta_1)]=(-1)^{\ell_1}\Theta_1(\theta_1)$ and $\mathbb{P}[\Theta_{K}(\theta_K)] = (-1)^{\ell_K-\ell_{K-1}}$ for $K>1$. 

\bibliography{adslib,otherlib}

\end{document}